\documentclass[twocolumn]{aastex63}

\usepackage{xspace}
\usepackage{enumitem}
\usepackage[utf8]{inputenc}
\usepackage{newunicodechar,graphicx}
\usepackage{amsmath}
\usepackage{lineno}
\usepackage{tabularx}
\usepackage{booktabs}
\usepackage{upgreek}
\usepackage{longtable}
\usepackage{footnote}
\usepackage[T1]{fontenc}
\usepackage{CJK}

\DeclareRobustCommand{\okina}{%
  \raisebox{\dimexpr\fontcharht\font`A-\height}{%
    \scalebox{0.8}{`}%
  }%
}
\newunicodechar{ʻ}{\okina}

\newcommand{\tess}{\textit{TESS}\xspace}

\newcommand{\mesa}{\textsf{MESA}\xspace}

\newcommand{\pone}{TOI-1181 b\xspace}
\newcommand{\ptwo}{TOI-6029 b\xspace}
\newcommand{\pthree}{TOI-4379 b\xspace}

\newcommand{\stwo}{TOI-6029\xspace}
\newcommand{\sthree}{TOI-4379\xspace}

\newcommand{\rpone}{$1.519\pm0.059$\xspace}
\newcommand{\rptwo}{$1.284\pm0.098$\xspace}
\newcommand{\rpthree}{$1.489\pm0.079$\xspace}

\newcommand{\mpone}{$1.179\pm0.023$\xspace}
\newcommand{\mptwo}{$1.635\pm0.032$\xspace}
\newcommand{\mpthree}{$1.113\pm0.071$\xspace}

\newcommand{\porbone}{$2.103189^{+0.000003}_{-0.000005}$\xspace}
\newcommand{\porbtwo}{$5.7987\pm0.0004$\xspace}
\newcommand{\porbthree}{$3.25352\pm0.00002$\xspace}

\newcommand{\lampone}{$-1.5^{+15.6}_{-15.3}$\xspace}
\newcommand{\lamptwo}{$-14.4^{+16.7}_{-12.7}$\xspace}
\newcommand{\lampthree}{$-1.0^{+22.1}_{-29.3}$\xspace}

\shorttitle{\tess GTG VI: Realigned Hot Jupiters}
\shortauthors{Saunders et al.}

\graphicspath{{./}{figures/}}

\begin{document}

\title{\tess Giants Transiting Giants VI: Newly Discovered Hot Jupiters Provide Evidence for Efficient Obliquity Damping After the Main Sequence}

\author[0000-0003-2657-3889]{Nicholas Saunders}
\altaffiliation{NSF Graduate Research Fellow}
\affiliation{Institute for Astronomy, University of Hawaiʻi at M\=anoa, 2680 Woodlawn Drive, Honolulu, HI 96822, USA}

\author[0000-0003-4976-9980]{Samuel K.\ Grunblatt}
\affiliation{Department of Physics and Astronomy, Johns Hopkins University, 3400 N Charles St, Baltimore, MD 21218, USA}

\author[0000-0003-2657-3889]{Ashley Chontos}
\altaffiliation{Henry Norris Russell Fellow}
\affiliation{Department of Astrophysical Sciences, Princeton University, 4 Ivy Lane, Princeton, NJ 08544, USA}

\author[0000-0003-2657-3889]{Fei Dai}
\affiliation{Institute for Astronomy, University of Hawaiʻi at M\=anoa, 2680 Woodlawn Drive, Honolulu, HI 96822, USA}

\author[0000-0001-8832-4488]{Daniel Huber}
\affiliation{Institute for Astronomy, University of Hawaiʻi at M\=anoa, 2680 Woodlawn Drive, Honolulu, HI 96822, USA}

\begin{CJK*}{UTF8}{gbsn}
\author[0000-0002-2696-2406]{Jingwen Zhang (张婧雯)}
\altaffiliation{NASA FINESST Fellow}
\affiliation{Institute for Astronomy, University of Hawaiʻi at M\=anoa, 2680 Woodlawn Drive, Honolulu, HI 96822, USA}

\author[0000-0001-7409-5688]{Gu{\dh}mundur Stef\'ansson}
\affiliation{Anton Pannekoek Institute for Astronomy, 904 Science Park, University of Amsterdam, Amsterdam, 1098 XH}

\author[0000-0002-4284-8638]{Jennifer L. van Saders}
\affiliation{Institute for Astronomy, University of Hawaiʻi at M\=anoa, 2680 Woodlawn Drive, Honolulu, HI 96822, USA}

\author[0000-0002-4265-047X]{Joshua N.\ Winn}
\affiliation{Department of Astrophysical Sciences, Princeton University, 4 Ivy Lane, Princeton, NJ 08544, USA}

\author[0000-0003-3244-5357]{Daniel Hey}
\affiliation{Institute for Astronomy, University of Hawaiʻi at M\=anoa, 2680 Woodlawn Drive, Honolulu, HI 96822, USA}


\author[0000-0001-8638-0320]{Andrew W. Howard}
\affiliation{Department of Astronomy, California Institute of Technology, Pasadena, CA 91125, USA}

\author[0000-0003-3504-5316]{Benjamin Fulton}
\affiliation{NASA Exoplanet Science Institute/Caltech-IPAC, MC 314-6, 1200 E. California Boulevard, Pasadena, CA 91125, USA}

\author[0000-0002-0531-1073]{Howard Isaacson}
\affiliation{Department of Astronomy, University of California Berkeley, Berkeley, CA 94720, USA}
\affiliation{Centre for Astrophysics, University of Southern Queensland, Toowoomba, QLD, Australia}

\author[0000-0001-7708-2364]{Corey Beard}
\altaffiliation{NASA FINESST Fellow}
\affiliation{Department of Physics \& Astronomy, University of California Irvine, Irvine, CA 92697, USA}

\author[0000-0002-8965-3969]{Steven Giacalone}
\altaffiliation{NSF Astronomy and Astrophysics Postdoctoral Fellow}
\affiliation{Department of Astronomy, California Institute of Technology, Pasadena, CA 91125, USA}

\author[0000-0002-4290-6826]{Judah van Zandt}
\affiliation{Department of Physics \& Astronomy, University of California Los Angeles, Los Angeles, CA 90095, USA}

\author[0000-0001-8898-8284]{Joseph M. Akana Murphey}
\altaffiliation{NSF Graduate Research Fellow}
\affiliation{Department of Astronomy and Astrophysics, University of California, Santa Cruz, CA 95064, USA}

\author[0000-0002-7670-670X]{Malena Rice}
\affiliation{Department of Astronomy, Yale University, New Haven, CT 06511, USA}

\author[0000-0002-3199-2888]{Sarah Blunt}
\affiliation{Department of Astronomy, California Institute of Technology, Pasadena, CA 91125, USA}
\affiliation{CIERA, Northwestern University, Evanston IL 60201, USA}

\author[0000-0002-1845-2617]{Emma Turtelboom}
\affiliation{Department of Astronomy, University of California Berkeley, Berkeley, CA 94720, USA}

\author[0000-0002-4297-5506]{Paul A. Dalba}
\affiliation{Department of Astronomy and Astrophysics, University of California, Santa Cruz, CA 95064, USA}

\author[0000-0001-8342-7736]{Jack Lubin}
\affiliation{Department of Physics \& Astronomy, University of California Irvine, Irvine, CA 92697, USA}

\author[0000-0002-4480-310X]{Casey Brinkman}
\affiliation{Institute for Astronomy, University of Hawaiʻi at M\=anoa, 2680 Woodlawn Drive, Honolulu, HI 96822, USA}

\author[0000-0003-3179-5320]{Emma M. Louden}
\affiliation{Department of Astronomy, Yale University, New Haven, CT 06511, USA}

\author[0000-0002-3221-3874]{Emma Page}
\affiliation{Department of Physics, Lehigh University, 16 Memorial Drive East, Bethlehem, PA 18015, USA}


\author[0000-0001-8621-6731]{Cristilyn N.\ Watkins}
\affiliation{Center for Astrophysics \textbar \ Harvard \& Smithsonian, 60 Garden Street, Cambridge, MA 02138, USA}

\author[0000-0001-6588-9574]{Karen A.\ Collins}
\affiliation{Center for Astrophysics \textbar \ Harvard \& Smithsonian, 60 Garden Street, Cambridge, MA 02138, USA}

\author[0000-0003-2163-1437]{Chris Stockdale}
\affiliation{Hazelwood Observatory, Australia}

\author[0000-0001-5603-6895]{Thiam-Guan Tan}
\affiliation{Perth Exoplanet Survey Telescope, Perth, Western Australia}

\author[0000-0001-8227-1020]{Richard P. Schwarz}
\affiliation{Center for Astrophysics \textbar \ Harvard \& Smithsonian, 60 Garden Street, Cambridge, MA 02138, USA}

\author[0000-0001-8879-7138]{Bob Massey}
\affiliation{Villa '39 Observatory, Landers, CA 92285, USA}

\author[0000-0002-2532-2853]{Steve B. Howell}
\affiliation{NASA Ames Research Center, Moffett Field, CA 94035, USA}

\author[0000-0001-7246-5438]{Andrew Vanderburg}
\affiliation{Department of Physics and Kavli Institute for Astrophysics and Space Research, Massachusetts Institute of Technology, Cambridge, MA 02139, USA}

\author{George R.\ Ricker}
\affiliation{Department of Physics, and Kavli Institute for Astrophysics and Space Research, Massachusetts Institute of Technology, 77 Massachusetts Ave., Cambridge, MA 02139, USA}

\author[0000-0002-4715-9460]{Jon M.\ Jenkins}
\affiliation{NASA Ames Research Center, Moffett Field, CA, 94035}



\author[0000-0002-6892-6948]{Sara Seager}
\affiliation{Department of Physics, and Kavli Institute for Astrophysics and Space Research, Massachusetts Institute of Technology, 77 Massachusetts Ave., Cambridge, MA 02139, USA}
\affiliation{Department of Earth, Atmospheric, and Planetary Sciences, Massachusetts Institute of Technology, 77 Massachusetts Ave., Cambridge, MA 02139, USA}
\affiliation{Department of Aeronautics and Astronautics, Massachusetts Institute of Technology, 77 Massachusetts Ave., Cambridge, MA 02139, USA}

\author[0000-0002-8035-4778]{Jessie L. Christiansen}
\affiliation{NASA Exoplanet Science Institute, IPAC, MS 100-22, Caltech, 1200 E. California Blvd., Pasadena, CA 91125, USA}

\author[0000-0002-6939-9211]{Tansu Daylan}
\affiliation{Department of Physics and McDonnell Center for the Space Sciences, Washington University, St. Louis, MO 63130, USA}


\author{Ben Falk}
\affiliation{Space Telescope Science Institute, 3700 San Martin Drive, Baltimore, MD 21218, USA}


\author[0009-0008-9808-0411]{Max Brodheim}
\affiliation{W. M. Keck Observatory, 65-1120 Mamalahoa Hwy., Kamuela, HI 96743}

\author[0009-0004-4454-6053]{Steven R. Gibson}
\affil{Caltech Optical Observatories, Pasadena, CA, 91125, USA}

\author[0000-0002-7648-9119]{Grant M. Hill}
\affiliation{W. M. Keck Observatory, 65-1120 Mamalahoa Hwy., Kamuela, HI 96743}

\author[0000-0002-6153-3076]{Bradford Holden}
\affiliation{UCO/Lick Observatory, Department of Astronomy and Astrophysics, University of California at Santa Cruz, Santa Cruz, CA 95064, USA}

\author[0000-0002-5812-3236]{Aaron Householder}
\affiliation{Department of Earth, Atmospheric and Planetary Sciences, Massachusetts Institute of Technology, Cambridge, MA 02139, USA}
\affil{Kavli Institute for Astrophysics and Space Research, Massachusetts Institute of Technology, Cambridge, MA 02139, USA}

\author{Stephen Kaye}
\affiliation{Caltech Optical Observatories, California Institute of Technology, Pasadena, CA 91125, USA}

\author[0000-0003-2451-5482]{Russ R. Laher}
\affiliation{NASA Exoplanet Science Institute/Caltech-IPAC, MC 314-6, 1200 E California Blvd, Pasadena, CA 91125, USA}

\author[0009-0004-0592-1850]{Kyle Lanclos}
\affiliation{W. M. Keck Observatory, 65-1120 Mamalahoa Hwy., Kamuela, HI 96743}

\author[0000-0003-0967-2893]{Erik A. Petigura}
\affiliation{Department of Physics \& Astronomy, University of California Los Angeles, Los Angeles, CA 90095, USA}

\author[0000-0001-8127-5775]{Arpita Roy}
\affiliation{Astrophysics \& Space Institute, Schmidt Sciences, New York, NY, 10011, USA}

\author[0000-0003-3856-3143]{Ryan A. Rubenzahl}
\altaffiliation{NSF Graduate Research Fellow}
\affiliation{Department of Astronomy, California Institute of Technology, Pasadena, CA 91125, USA}

\author[0000-0002-4046-987X]{Christian Schwab}
\affiliation{School of Mathematical and Physical Sciences, Macquarie University, Balaclava Road, North Ryde, NSW 2109, Australia}

\author[0000-0003-3133-6837]{Abby P. Shaum}
\affiliation{Department of Astronomy, California Institute of Technology, Pasadena, CA 91125, USA}

\author[0009-0007-8555-8060]{Martin M. Sirk}
\affiliation{Space Sciences Laboratory, University of California, 7 Gauss Way, Berkeley, CA 94720, USA}

\author{Christopher L. Smith}
\affiliation{Space Sciences Laboratory, University of California, 7 Gauss Way, Berkeley, CA 94720, USA}

\author[0000-0002-6092-8295]{Josh Walawender}
\affiliation{W. M. Keck Observatory, 65-1120 Mamalahoa Hwy., Kamuela, HI 96743}

\author[0000-0002-4037-3114]{Sherry Yeh}
\affiliation{W. M. Keck Observatory, 65-1120 Mamalahoa Hwy., Kamuela, HI 96743}



\begin{abstract}

    The degree of alignment between a star's spin axis and the orbital plane of its planets (the stellar obliquity) is related to interesting and poorly understood processes that occur during planet formation and evolution. Hot Jupiters orbiting hot stars ($\gtrsim$6250 K) display a wide range of obliquities, while similar planets orbiting cool stars are preferentially aligned. Tidal dissipation is expected to be more rapid in stars with thick convective envelopes, potentially explaining this trend. Evolved stars provide an opportunity to test the damping hypothesis, particularly stars that were hot on the main sequence and have since cooled and developed deep convective envelopes. We present the first systematic study of the obliquities of hot Jupiters orbiting subgiants that recently developed convective envelopes using Rossiter-McLaughlin observations. Our sample includes two newly discovered systems in the Giants Transiting Giants Survey (\ptwo, \pthree). We find that the orbits of hot Jupiters orbiting subgiants that have cooled below $\sim$6250 K are aligned or nearly aligned with the spin-axis of their host stars, indicating rapid tidal realignment after the emergence of a stellar convective envelope. We place an upper limit for the timescale of realignment for hot Jupiters orbiting subgiants at $\sim$500 Myr. Comparison with a simplified tidal evolution model shows that obliquity damping needs to be $\sim$4 orders of magnitude more efficient than orbital period decay to damp the obliquity without destroying the planet, which is consistent with recent predictions for tidal dissipation from inertial waves excited by hot Jupiters on misaligned orbits.
\\
\end{abstract}

\section{Introduction} \label{sec:intro}



Giant planets on close-in orbits (hot Jupiters) have been observed in a wide variety of stellar environments, but it remains unclear how they arrive at such small orbital separations from their host stars \citep{chatterjee2008,lee2017}. Two popular scenarios are disk-driven migration and high-eccentricity migration \citep{dawson2018}. In both theories, giant planets form at larger semi-major axes, before losing angular momentum to either the protoplanetary disk (in disk-driven migration; e.g. \citealt{lin96}) or to another planet or star (in high-eccentricity migration; e.g. \citealt{rasio1996}). In the latter case, the planet ends up on an eccentric orbit that gradually shrinks and circularizes as the eccentricity is damped by tidal interactions with the star. 

The stellar obliquity---the angle between the spin axis of the star and the planet's orbital axis---is a key piece of evidence to distinguish between these pathways as each scenario produces a distinct distribution of spin-orbit angles. Disk migration results in aligned populations at early times, while high-eccentricity migration predicts early scattering events that might alter the obliquity \citep{dawson2018}. Observations of hot Jupiters around main-sequence stars have shown that the stellar obliquity distribution has a strong dependence on the star's effective temperature \citep{winn2010,rice2022}, with hot hosts displaying a wide range of obliquities, and cool hosts being preferentially aligned. The critical separation is approximately at the Kraft break \citep{kraft1967}, around 6250 K, corresponding to the break between stars with primarily radiative envelopes ($>$6250 K) and stars with thick convective envelopes ($<$6250 K). The distinct distributions of obliquities separated by the Kraft Break suggest two possible scenarios:
\begin{enumerate}
    \item Hot Jupiters have unique formation pathways around hot and cool stars, resulting in aligned orbits around cool stars and a broad range of obliquities around hot stars.
    \item Hot Jupiters are formed by the same mechanism around hot and cool stars, and the spin-orbit angle is realigned during the main sequence by tidal dissipation in stars with convective envelopes.
\end{enumerate}

The correlation between spin-orbit alignment and the presence of a convective envelope has driven theories for realignment through the tidal damping of obliquity \citep{anderson2021}. During their main sequence lifetimes, low mass stars ($\lesssim$1.2 M$_\odot$) have surface convection zones that deepen as the star cools and evolves into a subgiant. However, it is challenging to constrain the timescales of damping mechanisms using a main sequence population in part due to uncertain age estimates. More massive stars $\gtrsim$1.2 M$_\odot$) have radiative envelopes on the main sequence, which cool and become convective as the stars drop below the Kraft break during the subgiant or red giant branch phase. As these stars only develop substantial convective envelopes after leaving the main sequence, they provide a unique avenue to test the timescales of tidal damping. If convection zones efficiently damp obliquities, as required by high-eccentricity migration, subgiants that were formerly 'hot' and had a broad range of obliquities should be found to have low obliquities due to rapid obliquity damping.

Studies have suggested that the presence of a convective envelope drives efficient damping of orbital period, eccentricity, and obliquity \citep{winn10,lai2012,rogers2013,albrecht2021,patel2022}. Diminishing orbital period has been unambiguously observed in only a small handful of systems \citep{maciejewski2016,maciejewski2018,yee2020,vissapragada2022}, though these measurements provide useful benchmarks for the timescales and efficiencies of mechanisms for orbital evolution. Orbital decay is driven by the transfer of orbital angular momentum to stellar rotation \citep{hut1980,hut1981}. As an orbiting planet raises a tidal bulge in the outer envelope of its host star, turbulent viscosity \citep{lai2012} and the excitation of waves through the stellar interior \citep{rogers2013} sap energy from the orbit. This orbital evolution can take place over human timescales \citep{yee2020,vissapragada2022}, and the changing orbital period can be observed.

The sky-projected obliquity can be constrained by obtaining high-resolution spectra during a planet's transit and modeling the Rossiter-McLaughlin (RM) effect \citep{rossiter1924,mclaughlin1924}. As a planet crosses in front of its host star, it blocks regions of the stellar surface that are red-shifted or blue-shifted relative to the observer due to the stellar rotation. These red- and blue-shifts imprint an additional signal onto the traditional radial velocity curve caused by the planet's orbit, the morphology and amplitude of which depend on the relative alignment of the planet's orbital path across the rotating star, and can be used to constrain the sky-projected obliquity of the system. The nearly all-sky coverage of NASA's Transiting Exoplanet Survey Satellite (\tess; \citealt{Ricker2015JATIS...1a4003R}) has enabled the discovery of a large sample of hot Jupiters orbiting bright subgiants which are amenable to measurements of the RM effect \citep{saunders2022,grunblatt2022,grunblatt2023,pereira2024}, allowing us measure spin-orbit alignment and understand late-stage planetary orbit evolution. 

In this paper we present observations of the RM effect for three hot Jupiters orbiting subgiants observed by \tess: \pone ($R_{\rm p} = $ \rpone $ R_{\rm J}$, $P_{\rm orb} =$ \porbone days, $R_\star = 1.89\pm0.18 R_\odot$, T$_{\rm eff} = 6106\pm100$ K), \ptwo ($R_{\rm p} = $ \rptwo $ R_{\rm J}$, $P_{\rm orb} =$ \porbtwo days, $R_\star = 2.23\pm0.05 R_\odot$, T$_{\rm eff} = 6223\pm100$ K), and \pthree ($R_{\rm p} = $ \rpthree $ R_{\rm J}$, $P_{\rm orb} =$ \porbthree days, $R_\star = 1.82\pm0.25 R_\odot$, T$_{\rm eff} = 6020\pm100$ K). \pone is a previously confirmed system \citep{kabath2022,chontos2024}, while \ptwo and \pthree are newly-identified planets confirmed by our work as part of the Giants Transiting Giants Survey. We find that all three systems have a low sky-projected obliquity ($|\lambda| < 15^\circ$), indicating that tidal realignment is efficient after the post-main-sequence emergence of a convective envelope. We then test models for realignment, and find that enhanced efficiency of obliquity damping after the main sequence can lead to realignment without inspiral and engulfment. 






\section{Observations} \label{sec:data}

\subsection{\tess Photometry} \label{sec:tess}


\ptwo and \pthree were discovered as part of our Giants Transiting Giants survey (GTG; \citealt{saunders2022,grunblatt2022,grunblatt2023,pereira2024}). \tess data are available for \ptwo in 30-minute cadence Full-Frame Images (FFIs) for Sectors 17, 18, and 24, spanning from October 7, 2019 - May 13, 2020, and in 200-second cadence FFIs for Sector 58 from October 29 - November 26, 2022. It was flagged by our survey as a planet candidate in 2021, and was selected for \tess 2-minute cadence observations in Sector 58. \tess data are available for \pthree in 30-minute FFIs in Sector 12 (April 22 - May 21, 2019), 10-minute cadence in Sector 39 (May 26 - June 24, 2021), and 200-second cadence in Sector 66 (June 2  - July 1, 2023). We identified \pthree as a planet candidate in 2021, and it received 2-minute cadence observations in Sectors 39 and 66. 

\pone was first identified by the MIT Quick Look Pipeline (QLP; \citealt{huang2020}) in 2019, and \tess data are available in 30-minute cadence FFIs for Sectors 14, 15, 17, 18, 19, 20, 21, 22, 23, 24, 25, and 26, spanning from July 18, 2019 - July 4, 2020. \tess data are available in 10-minute cadence FFIs for Sectors 40, 41, 47, 48, 49, 50, 51, 52, 53, 54, 55, spanning from June 24, 2021 - September 1, 2022. In 200-second cadence FFIs, it has so far been observed in Sectors 56, 57, 58, 59, and 60, covering September 1, 2022 - January 18, 2023. \pone was scheduled for 2-minute cadence observations beginning on April 16, 2020 with Sector 24, and was selected for 2-minute cadence data through Sector 60. The planet was confirmed by \cite{kabath2022}, and was followed up as part of the \tess-Keck Survey (TKS; \citealt{chontos2022,chontos2024}). 

All photometry data used are the Presearch Data Conditioning simple aperture photometry (PDC\_SAP; \citealt{smith12,stumpe12,stumpe2014}) \tess 2-minute light curves produced by the Science Processing Operations Center (SPOC; \citealt{jenkins16}) at NASA Ames Research Center downloaded from the Mikulski Archive for Space Telescopes (MAST).

We searched for additional transiting planets within these systems in the \tess photometry by masking the known transits and performing a Box-fitting Least Squares (BLS) periodogram analysis with the \textsf{lightkurve} implementation of the \textsf{astropy} BLS tools. We did not identify additional transiting planet signals.

The Science Processing Operations Center (SPOC; \citealt{jenkins16}) performed multi-transiting planet search in the subsequent 2-min cadence observations and detected the signatures of TOIs 1181 b, 6029 b and 4379 b using a noise-compensating matched filter \citep{jenkins2002,jenkins10,jenkins2020}. The transit signatures were fitted with an initial limb-darkened transit model \citep{li2019} and were subjected to a number of diagnostic tests \citep{twicken2018}, including the difference image centroiding test, which located the host stars within $0\arcsec.409$ $\pm$ $2\arcsec.6$, $0\arcsec.78$ $\pm$ $2\arcsec.5$, and $2\arcsec.6$ $\pm$ $2\arcsec.6$. The SPOC searches likewise failed to identify additional transiting planet signals.

\subsection{Radial Velocity Follow-up} \label{sec:RVs}

High precision RVs were taken using the High Resolution Echelle Spectrograph (HIRES) on the 10-meter Keck I telescope on Maunakea, Hawaiʻi \citep{vogt94}. For \pone, 87 RV measurements were taken between December 2, 2019 and May 16, 2022. For \ptwo, 46 RV measurements were taken between December 25, 2020 and October 6, 2022. For \pthree, 12 RV measurements were taken between September 21, 2021 and July 16, 2022. RV observations were taken in the HIRESr configuration with an iodine cell in the light path for wavelength calibration. The C2 decker was used for all observations, providing a spectral resolution of $R\approx60,000$. RVs were reduced with the California Planet Search (CPS; \citealt{howard10}) pipeline. Table \ref{tab:my_label} in Appendix \ref{sec:rv_obs} contains all RV observations and corresponding uncertainties used in this analysis.

\subsection{Rossiter-McLaughlin Observations} \label{sec:rm_obs}

In addition to RVs over the full phase of the planets' orbits, we obtained RV observations during each planet's transit to measure the Rossiter-McLaughlin (RM) effect. We obtained high resolution spectra during transit events with HIRES for \pone and \ptwo, and with the Keck Planet Finder (KPF; \citealt{kpf,kpf2018,kpf2020}) for \pthree. 

The HIRES observing configuration during the RM observations matches that described in \S \ref{sec:RVs}. We obtained 40 HIRES RV observations during the transit of \pone on September 8, 2020. The typical counts for each $\sim$10-minute exposure were 20,000 e$^-$ with an average RV uncertainty of $\sim$3 m s$^{-1}$. We obtained full coverage of the transit of \pone, with $\sim$1 hour before ingress and $\sim$1.5 hours after egress to establish an RV baseline. For \ptwo, we obtained 34 HIRES RVs on September 18, 2022 with typical counts of 5,000 e$^-$ over $\sim$13-minute exposures and average RV uncertainty of $\sim$5 m s$^{-1}$. Poor weather prevented observation of the full transit egress, and we were unable to continue the observations after the transit event concluded.

We obtained 30 observations of \pthree on June 6, 2023 using KPF, with 9-minute exposures. KPF is a fiber-fed spectrograph with a resolution of $R\approx98,000$. Observations were taken with simultaneous wavelength calibration from an etalon lamp fed through a dedicated calibration fiber. In addition to simultaneous calibrations, we took dedicated etalon calibration exposures (so-called ``slew-cals") each hour to monitor RV stability. Due to its southerly declination ($-32^\circ$), the window to observe the transit of \pthree was tight, and the star was not visible above the Keck telescope dome during ingress. RVs were measured using two distinct approaches---first, using a cross-correlation function (CCF) on a set of spectral lines produced by the public KPF data reduction pipeline\footnote{\href{https://github.com/Keck-DataReductionPipelines/KPF-Pipeline/}{https://github.com/Keck-DataReductionPipelines/KPF-Pipeline/}}. This reduction method resulted in a typical RV uncertainty of $\sim$7 m s$^{-1}$. We also reduced the RVs by analyzing the spectra using an adapted version of the SpEctrum Radial Velocity Analyser code (SERVAL; \citealt{zechmeister2018}), which measures shifts in the observed spectra relative to a matched template spectrum, but found that the CCF reduction resulted in RVs with higher precision and smaller scatter. The KPF data reduction pipeline is being actively developed, and future improvements to the measured RV precision are likely to be implemented.

\subsection{Ground-based Imaging} \label{sec:hci}

We obtained high-contrast imaging of \stwo and \sthree to search for close stellar companions which might have biased the measurement of the amplitude of the transit signal observed by \tess. \stwo was observed by the NIRC2 near-infrared imager in the Kp bandpass on the 10-meter Keck II telescope on June 26, 2023.

We first pre-processed the data using the Vortex Imaging Processing (VIP) software package \citep{GomezGonzalez2017, Christiaens2023}. We performed flat-fielding, bad-pixel removal, and correct for geometric distortions by applying the solution in \citet{Service2016}. Since the data were taken in vertical angle mode, we de-rotated each image according to the parallactic angles and then stacked the individual images into a combined image. To register the eight frames, we identify the star's position by fitting a 2D Gaussian to the stellar point spread function (PSF) in each frame.

A companion to \stwo was identified with a separation of $763 \pm 20$ milliarcseconds at a position angle of $110.2 \pm 0.5^{\circ}$. We measured a difference in magnitude in the Ks bandpass of $\Delta m=4.44 \pm 0.05$. The reduced and derotated image can be found in Figure \ref{fig:toi6029_image} in Appendix \ref{sec:imaging_figures}.

After obtaining the pre-processed cubes, we extracted the astrometry of the companion applying a least-squares minimization code. Specifically, we fit the PSF of the companion using the unsaturated PSF of the primary star in the same images (see details in \citealt{Xuan2024}).

\pthree was observed by the 8-meter Gemini South telescope on Cerro Pachon using the speckle interferometric instrument Zorro \citep{scott2021,howell2022} on May 18, 2022 UT. Three sets of 1000 $\times$ 60 ms images were obtained simultaneously at 562 and 832 nm, the data was reduced using our standard software pipeline \citep{howell11} and the limiting magnitude contrast was computed for each filter as a function of the angular separation from TOI-4379. The 5-σ contrast curves reached depths of 5 to 6.5 magnitudes in the two filters over the angular range of 0.1 to 1.2 arcsec. Within these contrast levels and angular limits, no close companion was detected. The raw and reduced data, including the contrast curves and reconstructed images at each wavelength, are available on the \textsf{EXOFOP-TESS} website\footnote{\href{exofop.ipac.caltech.edu/tess/target.php?id=97766057}{exofop.ipac.caltech.edu/tess/target.php?id=97766057}}.

\subsection{Ground-based Photometry\label{subsec:ground}}

The \textit{TESS} pixel scale is $\sim 21\arcsec$ pixel$^{-1}$, often causing multiple stars to blend in the \textit{TESS} photometric aperture. To confirm the true source of the \textit{TESS} detection, we acquired ground-based follow-up photometry of the fields around TOI-6029 and TOI-4379 as part of the \textit{TESS} Follow-up Observing Program \citep[TFOP;][]{collins:2019}\footnote{tess.mit.edu/followup}. We used the \textsf{TESS Transit Finder}, which is a customized version of the \textsf{Tapir} software package \citep{Jensen:2013}, to schedule our transit observations.

\subsubsection{LCOGT\label{subsubsec:lcogt}}

We observed an ingress window of \ptwo in Sloan $i'$ band on UTC 2023 August 03 from the Las Cumbres Observatory Global Telescope \citep[LCOGT;][]{Brown:2013} 1\,m network node at McDonald Observatory near Fort Davis, Texas, United States (McD). The 1\,m telescopes are equipped with $4096\times4096$ SINISTRO cameras having an image scale of $0\farcs389$ per pixel, resulting in a $26\arcmin\times26\arcmin$ field of view. The images were calibrated by the standard LCOGT \textsf{BANZAI} pipeline \citep{McCully:2018}, and differential photometric data were extracted using \textsf{AstroImageJ} \citep{Collins:2017}. The transit is detected within the $3\farcs5$ follow-up photometric aperture which excludes the flux from the nearest known neighbor in the Gaia DR3 catalog (Gaia DR3 430269136521931776) that is $\sim17\arcsec$ north of \stwo.

We also observed a full transit window of \pthree in Sloan $i'$ band on May 15, 2022 (UTC) from the LCOGT 1\,m network node at Cerro Tololo Inter-American Observatory in Chile (CTIO). The transit is detected within the $2\farcs7$ follow-up photometric aperture, which excludes most of the flux from the nearest known neighbor in the Gaia DR3 catalog (Gaia DR3 6027376203516633728), which is $\sim3\arcsec$ west of \sthree. 

\subsubsection{PEST}
We observed a full transit window of \pthree in Sloan $g'$ band on May 27, 2023 (UTC) using the Perth Exoplanet Survey Telescope (PEST), located near Perth, Australia. The 0.3 m telescope is equipped with a $5544\times3694$ QHY183M camera with an image scale of 0$\farcs$7 pixel$^{-1}$, resulting in a $32\arcmin\times21\arcmin$ field of view. A custom pipeline based on \textsf{C-Munipack}\footnote{c-munipack.sourceforge.net} was used to calibrate the images and extract the differential photometry. The transit is detected within the $6\farcs4$ follow-up photometric aperture, which includes the flux from the nearest known neighbor in the Gaia DR3 catalog. Figure \ref{fig:ground_phot} shows the ground-based transit observations of \ptwo and \pthree with best-fit transit models. All light curve data are available on the \textsf{EXOFOP-TESS} website\footnote{\href{exofop.ipac.caltech.edu/tess/target.php?id=419523962}{exofop.ipac.caltech.edu/tess/target.php?id=419523962}}, but are not included in the joint model of this work. 

\section{Host Star Characterization} \label{sec:host_stars}

\subsection{High-resolution Spectroscopy} \label{sec:stellar_params}

We obtained out-of-transit, iodine-free spectra with Keck/HIRES for each of our host stars to extract precise stellar parameters. Stellar properties were computed using SpecMatch empirical (SpecMatch-Emp; \citealt{yee17}) and synthetic (SpecMatch-Synth; \citealt{petigura15}) methodologies. All of our host stars have effective temperatures T$_{\rm eff} > 5000$ K, and we therefore adopt values from SpecMatch-Synth. Our adopted values can be found in Table \ref{tab:stellar}.

\subsection{Isochrone Fitting} \label{sec:isoclassify}

We used the \textsf{isoclassify}\footnote{\href{https://github.com/danxhuber/isoclassify}{https://github.com/danxhuber/isoclassify}} Python package \citep{huber2017,berger20} to compute additional stellar properties. We ran the code in grid mode where we pass input observables taken from SpecMatch-Synth and position and parallax distance from Gaia DR3 \citep{gaiadr3} and K-band magnitude adopted from the \tess Input Catalog (TIC; \citealt{stassun2019,paegert2022} v8.2. \textsf{isoclassify} uses the stellar model grid to compute stellar mass ($M_\star$), stellar radius ($R_\star$), surface gravity ($\log{g}$), distance ($d$), luminosity ($L$), and stellar age. We used the Combined19 allsky dust map from \textsf{mwdust} \citep{bovy16}. 

We account for the companion to \stwo described in \S \ref{sec:hci} by running \textsf{isoclassify} in binary mode, including the correction for the measured companion magnitude. For the companion to \stwo, we obtain a mass of $0.44\pm0.02$ M$_\odot$, radius of $0.45\pm0.02$ R$_\odot$, and temperature of $3528\pm40$ K.

The \textsf{isoclassify} outputs and associated uncertainties can be found in Table \ref{tab:stellar}. All host stars are found to be evolved ($\sim$2 R$_\odot$) stars with masses between 1.2 and 1.6 M$_\odot$ and metallicities between 0.1 and 0.4 dex.

\subsection{Asteroseismology} \label{sec:asteroseismology}

Finally, to improve the characterisation of the host stars, we searched for signatures of stellar p-mode oscillations in the light curve.  To do this, we first remove transits and apply a simple high pass-filter to remove further low-frequency variability. We use the PySYD pipeline \citep{Chontos2021textttpySYD} to search for variability in the host stars. We find no evidence of oscillations in TOI-6029 and TOI-4379. In TOI-1181, however, we identify a weak signal associated with the global oscillation power excess, commonly referred to as $\nu_{\rm max}$, at 737$\pm$69 $\mu$Hz at an SNR of 4.55. This value is slightly lower than the expected $\nu_{\rm max}$ of 930~$\mu$Hz based on asteroseismic scaling relations (cf. \citealt{stello09}). As a result of this, and non-detections in the other host stars, we do not use asteroseismology to constrain the stellar parameters.


\begin{table*}[]
\footnotesize
    \begin{tabular}{l l c c c c}
    \toprule
         & & \textbf{\pone} & \textbf{\ptwo} & \textbf{\pthree} & Source \\
         \midrule
         TIC ID &  & 229510866 & 419523962 & 97766057 & a \\
         RA & & $19:48:51.8$ & $00:33:32.7$ & $16:48:11.49$ & a \\
         dec & & $+64:21:15.66$ & $+61:51:38.76$ & $-32:26:01.51$ & a \\
         V Mag & & $10.582\pm0.006$ & $12.858\pm0.038$ & $12.661\pm0.092$ & a \\
         \textit{Gaia} Mag & & $10.4776\pm0.0003$ & $12.2681\pm0.0002$ & $12.4083\pm0.0002$ & a \\
         \textit{TESS} Mag & & $10.079\pm0.0086$ & $11.6593\pm0.006$ & $11.8994\pm0.0069$ & a \\
         T$_{\rm eff}$ & (K) & $6054\pm81$ & $6223\pm100$ & $6020\pm100$ & b* \\ \relax
         [Fe/H] & (dex) & $0.38\pm0.06$ & $0.37\pm0.06$ & $0.16\pm0.06$ & b* \\
         $v\sin{i_\star}$ & (km s$^{-1}$) & $10.6\pm1.0$ & $10.4\pm1.0$ & $5.9\pm1.0$ & b* \\
         M$_\star$ & (M$_\odot$) & $1.467\pm0.024$ & $1.548^{+0.027}_{-0.085}$ & $1.372^{+0.037}_{-0.105}$ & c* \\
         R$_\star$ & (R$_\odot$) & $1.961\pm0.030$ & $2.338^{+0.027}_{-0.085}$ & $1.976\pm0.040$ & c* \\
         $\log{g}$ & (dex) & $4.02\pm0.02$ & $3.886^{+0.016}_{-0.024}$ & $3.929^{+0.038}_{-0.023}$ & c* \\
         Age & (Gyr) & $2.7\pm0.2$ & $2.31^{+0.0.58}_{-0.19}$ & $4.51^{+0.40}_{-1.21}$ & c* \\
         L & (L$_\odot$) & $4.39^{+0.23}_{-0.15}$ & $6.30^{+0.27}_{-0.30}$ & $4.34^{+0.28}_{-0.25}$ & c \\
         $d$ & (pc) & $310.5^{+6.2}_{-4.8}$ & $596.7^{+11.5}_{-11.1}$ & $617.4^{+11.6}_{-11.9}$ & c \\
         \bottomrule
    \end{tabular}
    \caption{Stellar properties derived for each of the host stars in our sample. Sources: (a) \textit{TESS} input catalog \citep{stassun2019}, (b) SpecMatch-Synth (this work; \citealt{petigura15}), (c) \textsf{isoclassify} (this work; \citealt{huber2017,berger20}). *Stellar properties for TOI-1181 adopted from Chontos et al. (in prep).}
    \label{tab:stellar}
\end{table*}

\section{Planet Modeling} \label{sec:modeling}

\subsection{Joint Transit and RV Fit} \label{sec:transit_rv_fit}

We used the \textsf{allesfitter} Python package \citep{allesfitter-paper,allesfitter-code} to simultaneously model the planet transit, radial velocity curve, and RM effect. All photometry data used are \tess 2-minute cadence observations reduced by the Science Processing Operations Center (SPOC; \citealt{spoc}) pipeline. We used the 2-minute cadence data as it provides the best constraint on the ingress and egress times, which are important for characterizing the shape of the Rossiter-McLaughlin signal. These data are also reduced with a uniform approach which accounts for contamination by nearby stars, providing the most accurate constraint on the transit depth and planet radius. To enable longer sampling within reasonable computation time ($\lesssim 24$ hours), we select only one Sector of 2-minute cadence data for each target (Sectors 73, 58, 66 for \pone, \ptwo, and \pthree, respectively). Including additional Sectors did not significantly impact the measured transit depth, period, or epoch.

We fit for orbital period, $P$, transit epoch, $t_0$, ratio of planetary to stellar radius, $R_p/R_\star$, cosine of orbital inclination, $\cos{i}$, RV semi-amplitude, $K$, sky-projected spin-orbit angle, $\lambda$, and the eccentricity and argument of periastron, parameterized by the expressions $\sqrt{e}\cos{\omega}$ and $\sqrt{e}\sin{\omega}$. For each instrument, we also fit for two quadratic limb darkening coefficients, described by $q_{1,{\rm TESS}}$, $q_{2,{\rm TESS}}$, $q_{1,{\rm HIRES}}$, $q_{2,{\rm HIRES}}$, $q_{1,{\rm KPF}}$, and $q_{2,{\rm KPF}}$. We include an offset parameter for the RV instruments, which captures small global offsets for all RVs taken by a given instrument. 


For each target, we tested (a) constant instrumental offset shared between all observations for a given instrument and (b) instrumental offset with a linear slope to capture long term trends in the RV signal. The linear offset improved the fit by reducing structure in the residuals for \ptwo, while the fits to \pone and \pthree preferred the constant intrumental offset. We predicted the RV signal due to the resolved companion for \ptwo to test whether it could cause the observed trend, assuming the companion was observed at maximum separation on a circular orbit. We found that the maximum slope caused by a companion of mass M$=0.44$ M$_\odot$ at the observed angular separation is $0.001$ m s$^{-1}$ day$^{-1}$, a factor of 10 smaller than the observed slope of $\sim0.01$ m s$^{-1}$ day$^{-1}$. We conclude that the outer companion is unlikely to be the source of the RV slope and may indicate the presence of an unresolved stellar or planetary-mass companion. Further monitoring of RV and astrometry for this system is required to determine the source of the RV trend.


We also included the RM effect in our \textsf{allesfitter} model (see \citealt{allesfitter-code} for details). The amplitude of the RM effect depends on the ratio of the planet's radius ($R_{\rm p}$) to the star's radius ($R_\star$), the projected rotation velocity of the star ($v\sin{i}$), and impact parameter ($b$), given by \cite{triaud2018} as:
\[
A_{\rm RM} \approx \frac{2}{3} \left(\frac{R_{\rm p}}{R_\star}\right)^2 v\sin{i}\sqrt{1-b^2}.
\]
As stars evolve off of the main sequence, they become larger, more luminous, and slower in their rotation. These changes have three primary effects that make RM observations challenging---first, the larger radius causes the transit duration to increase, limiting available observation windows; second, the increased stellar radius also causes the ratio $R_{\rm p}/R_\star$ to decrease, reducing the RM amplitude; finally, the slowed stellar rotation ($v\sin{i_\star}$) also reduces the RM amplitude. Additionally, granulation for subgiants can introduce RV jitter with an amplitude of $\sim$5 m s$^{-1}$ on timescales of hours \citep{tayar19}. 

By modeling the RM signal jointly with the photometric and RV observations, the constraints on the obliquity are strengthened. The RVs provide additional information about the out-of-transit RV baseline, and the deviation from that baseline informs the spin-orbit angle. The photometric transit gives a constraint on the planet radius, which plays a strong role in the amplitude of the RM signal. In turn, the amplitude of the RM signal sheds additional light on the planet radius. Because we have an independent constraint on the planet radius, the projected stellar rotation velocity, $v\sin{i_\star}$, is also well-constrained by the amplitude of the RM signal. 

\begin{figure*}[ht!]
    \centering
    \includegraphics[width=1\textwidth]{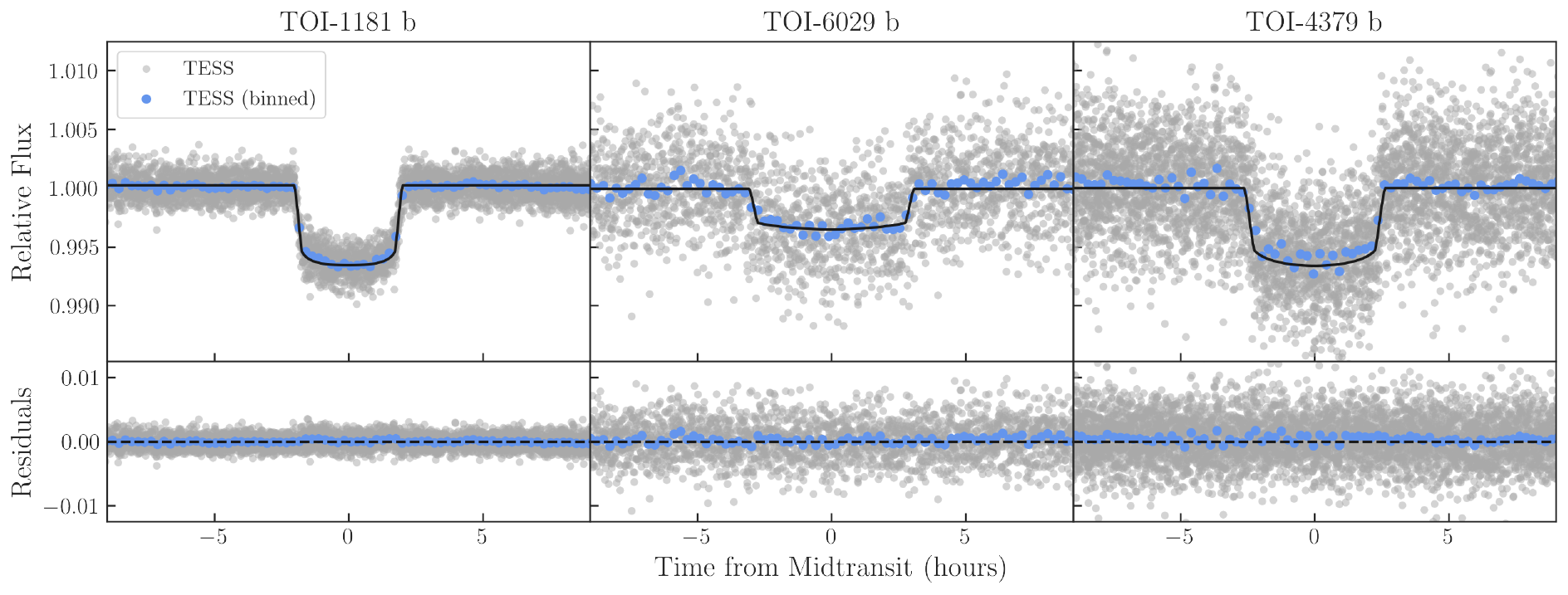}
    \caption{\textit{Top:} Normalized \tess photometry for each of our systems, phase-folded to the orbital periods of the planets. Raw \tess photometry is marked by gray points, and binned photometry is marked by blue points. The median of the posterior transit model is shown by the black line. \textit{Bottom:} Residuals between observations and the best-fit transit model.}
    \label{fig:tess_lcs}
\end{figure*}
\begin{figure*}[ht!]
    \centering
    \includegraphics[width=1\textwidth]{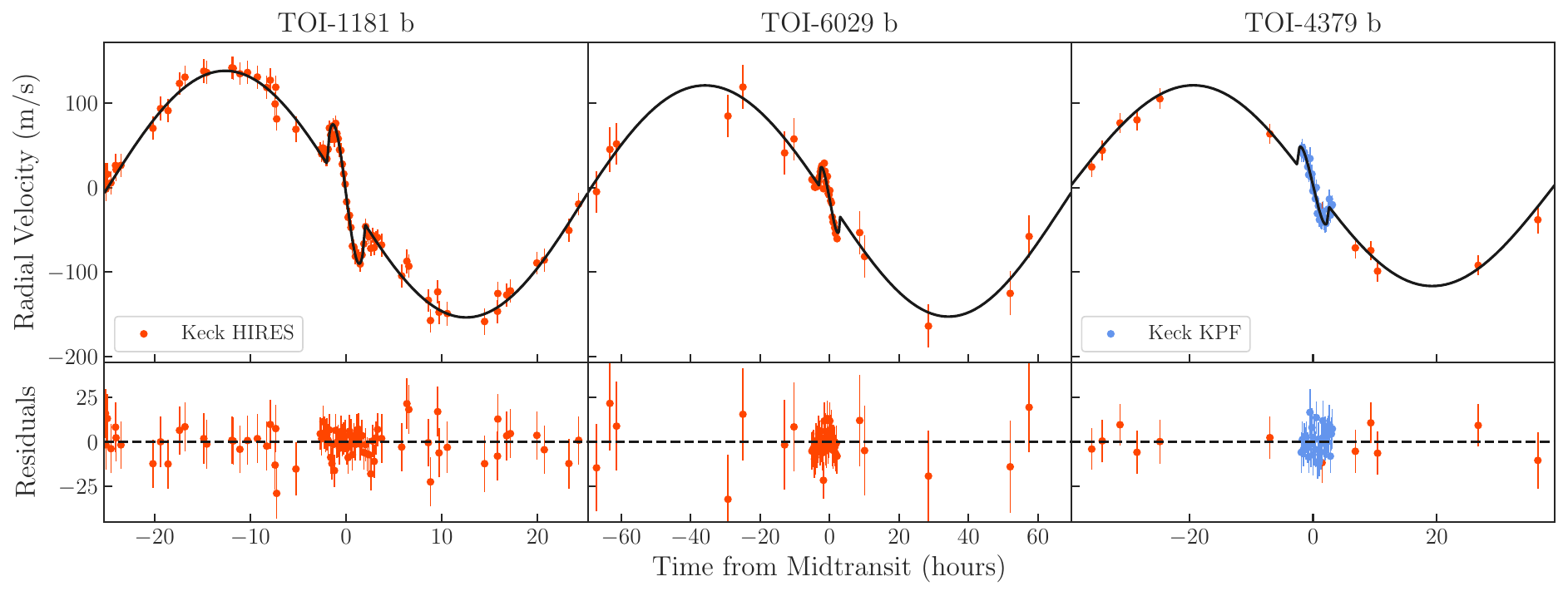}
    \caption{\textit{Top:} Radial velocity measurements for each of our systems, phase-folded to the orbital periods of the planets. Keck/HIRES observations are shown as red points and Keck/KPF observations are marked by blue points. The black line shows the best-fit orbital model to the system, including the Rossiter-McLaughlin signal. \textit{Bottom:} Residuals between observations and the best-fit RV model.}
    \label{fig:rvs}
\end{figure*}
\begin{figure*}[ht!]
    \centering
    \includegraphics[width=1\textwidth]{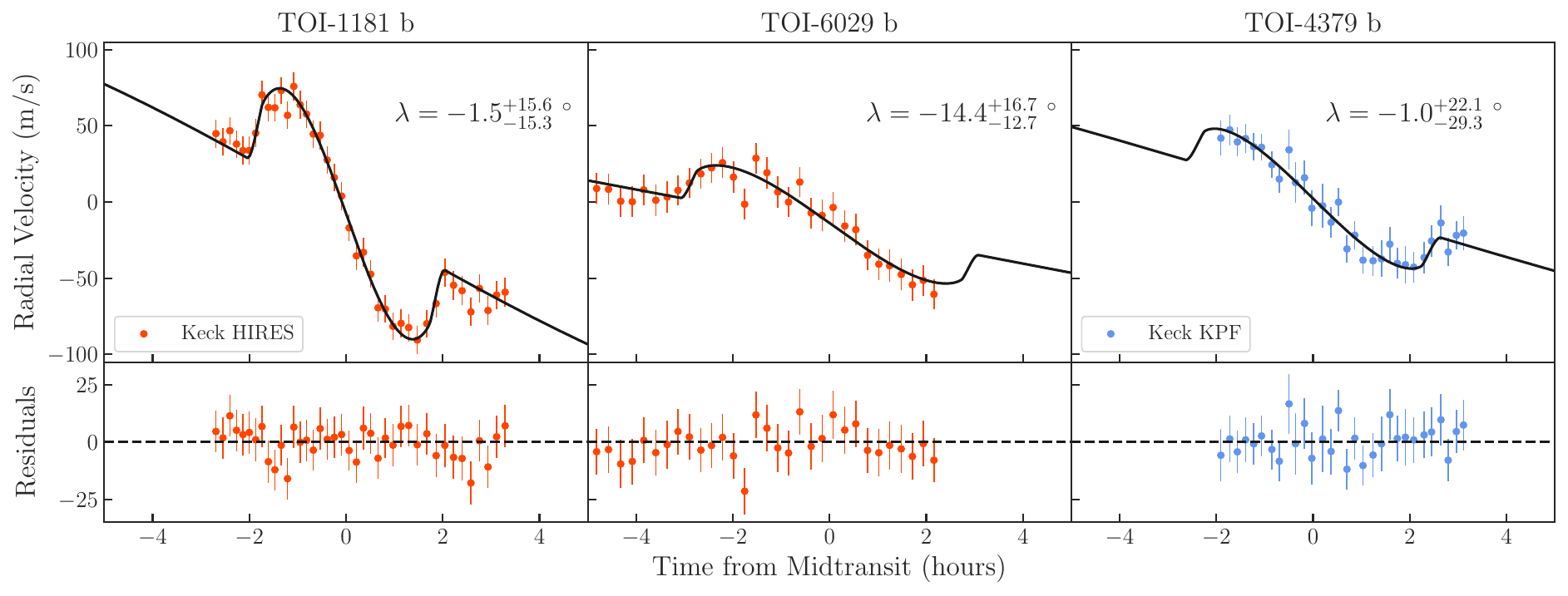}
\caption{\textit{Top:} Observations of the Rossiter-McLaughlin effect for each of our targets. Data from Keck/HIRES and Keck/KPF are shown as red and blue points, respectively, and the line shows our model fit. \textit{Bottom:} Residuals between observations and the best-fit RM model.}
\label{fig:rm}
\end{figure*}
\begin{figure}[h!]
    \centering
    \includegraphics[width=.475\textwidth]{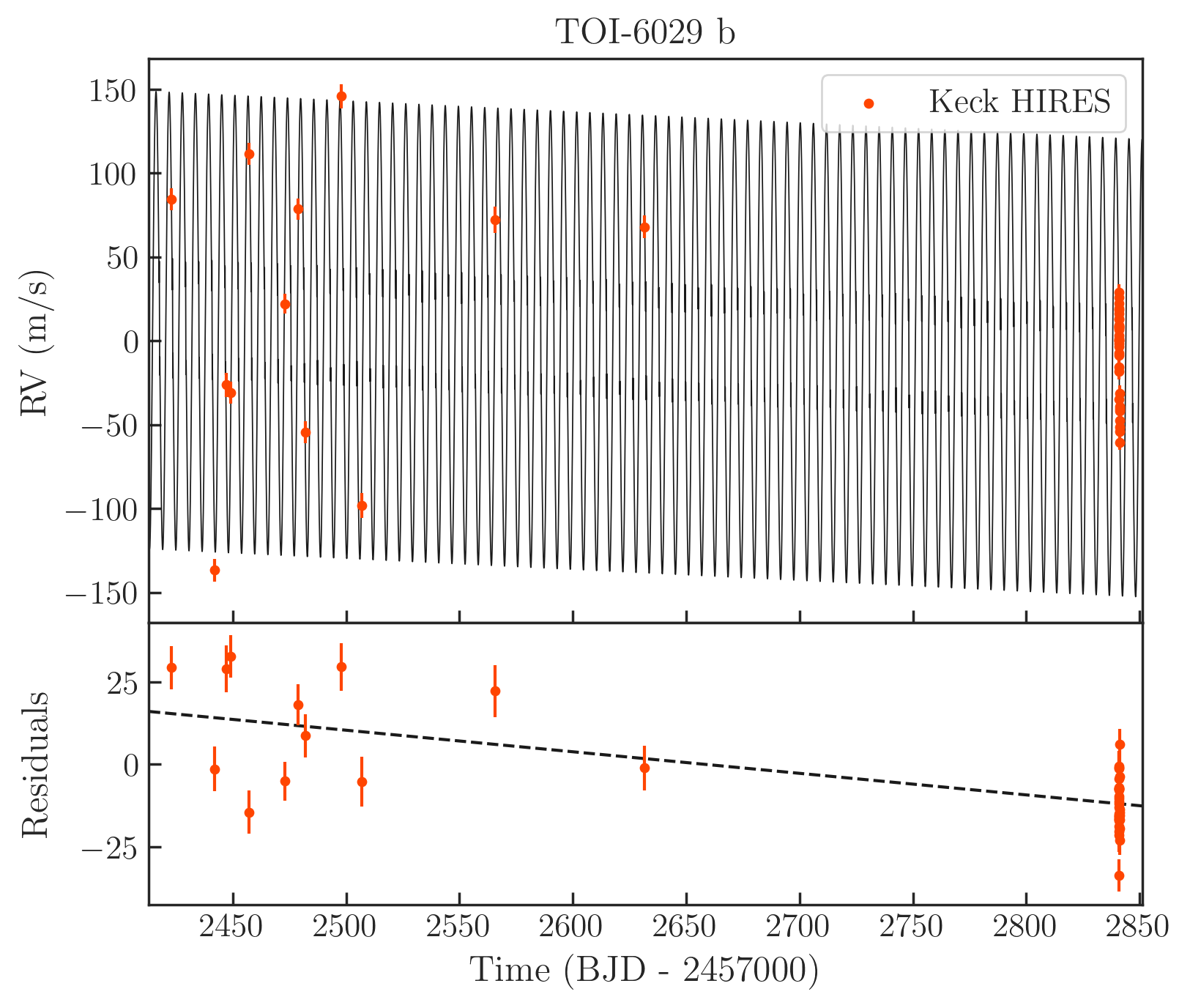}
    \caption{\textit{Top:} RV observations for TOI-6029 b, shown in red, with the full RV model shown in black. \textit{Bottom:} Residuals after removing the RV contribution from the transiting planet.}
    \label{fig:toi6029_rv}
\end{figure}

All priors and inferred values for our model parameters can be found in Table \ref{tab:fit_results}. For each planet, we additionally derive planet mass, $M_p$, planet radius, $R_p$, and eccentricity, $e$ from the outputs of our model. Figures \ref{fig:tess_lcs}, \ref{fig:rvs}, and \ref{fig:rm} show the transit, RV, and RM fits, respectively. For \ptwo, a significant long-term RV trend was detected, shown in Figure \ref{fig:toi6029_rv}. 


For each target, we sample the posterior distribution using affine-invariant Markov Chain Monte Carlo (MCMC). We initiate 100 walkers with 50,000 sampling steps and 5,000 burn-in steps per walker, ensuring that the chains ran at least 30 times the autocorrelation length and successfully converged. The residual scatter in the RVs is consistent with what we would expect for a subgiant ($10-15$ m s$^{-1}$), and there is no remaining structure in the residual \tess photometry.

\begin{table*}[]
\scriptsize
    \begin{tabular}{l l c c c}
    \toprule
         & & \textbf{\pone} & \textbf{\ptwo} & \textbf{\pthree} \\
         \midrule
         \multicolumn{3}{l}{\textit{Fitted parameters:}} & & \\
$P$ & (days) & \porbone & \porbtwo & \porbthree \\
$t_0$ & (BJD - 2457000) & $2481.51696^{+0.00055}_{-0.00057}$ & $2667.067^{+0.015}_{-0.017}$ & $2734.269^{+0.048}_{-0.003}$ \\
$R_{\rm p}/R_\star$ &  & $0.07947^{+0.00042}_{-0.00045}$ & $0.056^{+0.00074}_{-0.00074}$ & $0.077^{+0.002}_{-0.002}$ \\
$a / R_\star$ & & $3.976^{+0.078}_{-0.049}$ & $7.17^{+0.39}_{-0.31}$ & $4.63^{+0.56}_{-0.21}$  \\
$\cos{i}$ &  & $0.018^{+0.028}_{-0.013}$ & $0.023^{+0.020}_{-0.015}$ & $0.041^{+0.049}_{-0.029}$ \\
$K$ & (m s$^{-1}$) & $146\pm2$ & $137^{+9}_{-10}$ & $119^{+8}_{-5}$ \\
$v\sin{i}$ & (km s$^{-1}$) & $12.96^{+0.88}_{-0.64}$ & $11.38^{+0.45}_{-0.76}$ & $6.41^{+0.42}_{-0.66}$ \\
$\lambda$ & (deg) & $-1.5^{+15.9}_{-15.3}$ & $-14.4^{+16.7}_{-12.7}$ & $-1.0^{+22.1}_{-29.3}$ \\
$\sqrt{e}\cos{\omega}$ &  & $-0.040^{+0.056}_{-0.046}$ & $-0.130^{+0.120}_{-0.083}$ & $-0.018^{+0.088}_{-0.083}$ \\
$\sqrt{e}\sin{\omega}$ &  & $0.042^{+0.079}_{-0.086}$ & $-0.040^{+0.164}_{-0.148}$ & $0.096^{+0.159}_{-0.140}$ \\
$q_{1,{\rm TESS}}$ &  & $0.119^{+0.032}_{-0.014}$ & $0.041^{+0.030}_{-0.019}$ & $0.114^{+0.197}_{-0.074}$ \\
$q_{2,{\rm TESS}}$ &  & $0.156^{+0.078}_{-0.041}$ & $0.767^{+0.168}_{-0.269}$ & $0.404^{+0.376}_{-0.283}$ \\
$q_{1,{\rm HIRES}}$ &  & $\ldots$ & $\ldots$ & $\ldots$ \\
$q_{2,{\rm HIRES}}$ &  & $\ldots$ & $\ldots$ & $\ldots$ \\
$q_{1,{\rm KPF}}$ &  & $\ldots$ & $\ldots$ & $0.620^{+0.263}_{-0.327}$ \\
$q_{2,{\rm KPF}}$ &  & $\ldots$ & $\ldots$ & $0.589^{+0.287}_{-0.359}$ \\
\multicolumn{3}{l}{\textit{Derived parameters:}} & & \\
$M_p$ & ($M_J$) & \mpone & \mptwo & \mpthree \\
$R_p$ &($R_J$) & \rpone & \rptwo & \rpthree \\
$e$ &  & $0.011^{+0.007}_{-0.008}$ & $0.091^{+0.017}_{-0.015}$ & $0.018^{+0.029}_{-0.018}$ \\
\midrule 
\multicolumn{3}{l}{\textit{\textsf{Allesfitter} priors:}} & & \\
$P$ & (days) & $\mathcal{U}[2.1032; 2.0, 2.2]$ & $\mathcal{U}[5.7998; 5.75, 5.85]$ & $\mathcal{U}[3.2535;3.2, 3.3]$ \\
$t_0$ & (BJD - 2457000) & $\mathcal{U}[1684.4058; 1684.0, 1685.0]$ & $\mathcal{U}[2307.4306; 2307.0, 2308.0]$ & $\mathcal{U}[1628.0701;1628.0, 1628.1]$ \\
$R_{\rm p}/R_\star$ & & $\mathcal{U}[0.08; 0.01, 1.0]$ & $\mathcal{U}[0.057; 0.0, 1.0]$ & $\mathcal{U}[0.075;0.05, 1.0]$ \\
$\cos{i}$ & & $\mathcal{U}[0.01; 0.0, 1.0]$ & $\mathcal{U}[0.01; 0.0, 1.0]$ & $\mathcal{U}[0.01;0.0,1.0]$ \\
$K$ & (m s$^{-1}$) & $\mathcal{U}[150; 50, 400]$ & $\mathcal{U}[150; 0, 400]$ & $\mathcal{U}[120;50,400]$ \\
$v\sin{i_\star}$ & (km s$^{-1}$) & $\mathcal{U}[10.71;5.0,15.0]$ & $\mathcal{U}[10.4; 9.0, 12.0]$ & $\mathcal{U}[5.9;4.0,7.0]$ \\
$\lambda$ & (deg) & $\mathcal{U}[0.0;-180.0,180.0]$ & $\mathcal{U}[0.0;-180.0,180.0]$ & $\mathcal{U}[0.0;-180.0,180.0]$ \\
$\sqrt{e}\cos{\omega}$ & & $\mathcal{U}[0,0;0.0,1.0]$ & $\mathcal{U}[0.0;0.0,1.0]$ & $\mathcal{U}[0.0;0.0,1.0]$ \\
$\sqrt{e}\sin{\omega}$ & & $\mathcal{U}[0.0;0.0,1.0]$ & $\mathcal{U}[0.0;0.0,1.0]$ & $\mathcal{U}[0.0;0.0,1.0]$ \\
$q_{1,{\rm TESS}}$ & & $\mathcal{U}[0.5; 0.0, 1.0]$ & $\mathcal{U}[0.5; 0.0, 1.0]$ & $\mathcal{U}[0.5; 0.0, 1.0]$ \\
$q_{2,{\rm TESS}}$ & & $\mathcal{U}[0.5; 0.0, 1.0]$ & $\mathcal{U}[0.5; 0.0, 1.0]$ & $\mathcal{U}[0.5; 0.0, 1.0]$ \\
$q_{1,{\rm HIRES}}$ & & $\mathcal{U}[0.5; 0.0, 1.0]$ & $\mathcal{U}[0.5; 0.0, 1.0]$ & $\mathcal{U}[0.5; 0.0, 1.0]$ \\
$q_{2,{\rm HIRES}}$ & & $\mathcal{U}[0.5; 0.0, 1.0]$ & $\mathcal{U}[0.5; 0.0, 1.0]$ & $\mathcal{U}[0.5; 0.0, 1.0]$\\
$q_{1,{\rm KPF}}$ & & \ldots & \ldots & $\mathcal{U}[0.5; 0.0, 1.0]$ \\
$q_{2,{\rm KPF}}$ & & \ldots & \ldots & $\mathcal{U}[0.5; 0.0, 1.0]$ \\
\bottomrule
    \end{tabular}
    \caption{Fit and derived planet parameters, and \textsf{allesfitter} model priors for each of our systems. $\mathcal{U}[a;b,c]$ denotes a uniform distribution from $b$ to $c$, starting with the test value $a$.}
    \label{tab:fit_results}
\end{table*}

\subsection{Doppler Shadow} \label{sec:dt}

\begin{figure}[ht!]
    \centering
    \includegraphics[width=.5\textwidth]{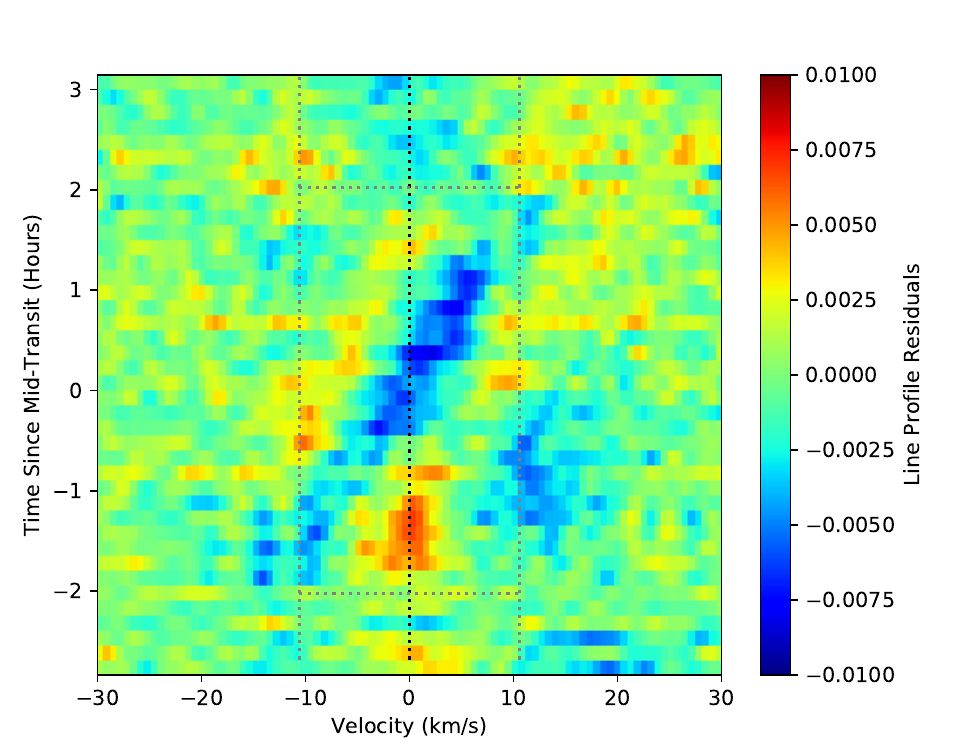}
\caption{Measurement of the Doppler shadow signal for TOI-1181 b from line profile residuals in the HIRES spectra. The horizontal dotted line marks the ingress and egress of the transit, the vertical dotted line indicate approximately the rotational broadening of the star ($v\sin{i}$). The planetary shadow is a diagonal pattern moving from the blue-shifted to the red-shifted part of the stellar disk. This is consistent with a well-aligned orbit seen in the RM measurement.}
\label{fig:dt_1181}
\end{figure}

\begin{figure}[h!]
    \centering
    \includegraphics[width=.5\textwidth]{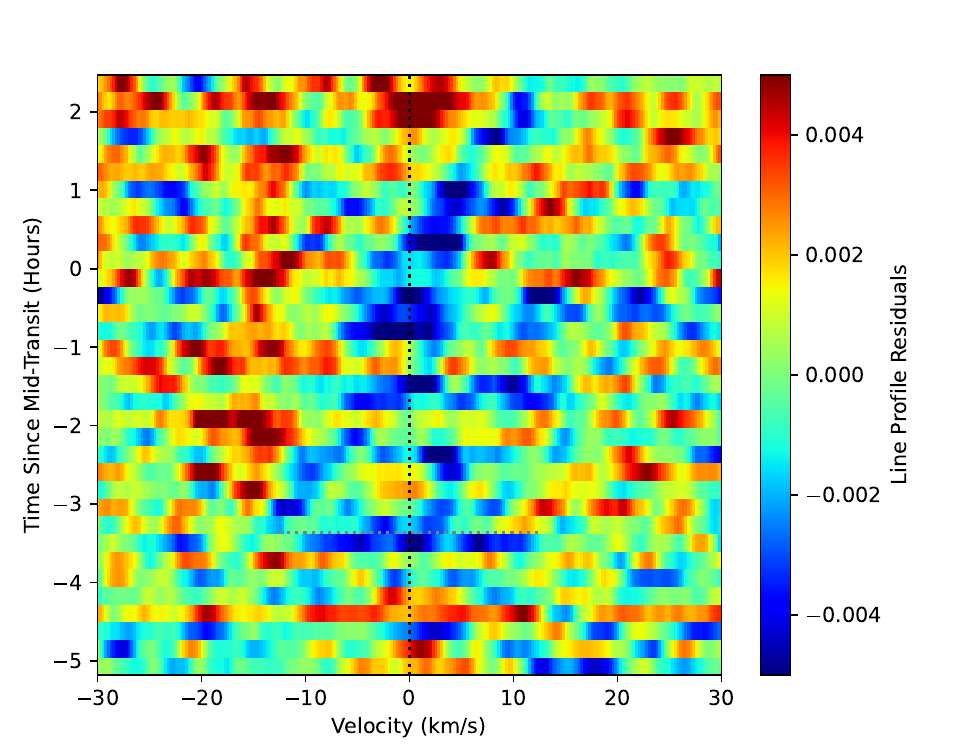}
    \caption{Same as Figure \ref{fig:dt_1181} but for \ptwo.}
    \label{fig:dt_6029}
\end{figure}

As an independent measure of the obliquity, we analyzed the spectral timeseries for each of our systems, and identified a Doppler shadow signal in the HIRES observations for \pone and \ptwo (Figures \ref{fig:dt_1181} and \ref{fig:dt_6029}, respectively). We selected lines in the non-Iodine portion of the HIRES spectra (4000-5000 \AA) to search for variations in the line profile during transit. We performed 5-$\sigma$ outlier clipping and removed the continuum and blaze function with a polynomial fit. Each spectrum was then cross-correlated with the SpecMatch template spectrum and the globabally-averaged line profile was subtracted off to isolate line profile variations. Our approach follows \cite{dai2020}. Analysis of the KPF data for \pthree did not result in the Doppler shadow detection, due to low signal-to-noise ratio of the observations, and the fact that the line profile is dominated by turbulence rather than rotational modulation.

\section{Results} \label{sec:results}

\subsection{TOI-1181} \label{sec:toi1181_results}

\pone is an intermediate-mass ($M_p=$ \mpone $ M_J$) hot Jupiter on a circular orbit. The planet radius is typical for a hot Jupiter of this mass ($R_p=$ \rpone $ R_J$). At a T$_{\rm eff}$ of 6054 K, its subgiant host star has cooled after its main sequence lifetime and is likely to have developed a non-negligible surface convective zone. The RM effect of \pone has the highest amplitude in our sample, due in part to the high projected rotation velocity of its host star ($v\sin{i_\star} = 10.6\pm1.0$ km s$^{-1}$), providing the most precise measurement of the spin-orbit angle. The posterior from our model gives the constraint on the sky-projected obliquity of TOI-1181 of $\lambda =$ \lampone $^\circ$. For \pone, we measure an obliquity from the Doppler shadow analysis of $7.5\pm17.2^\circ$, which is consistent with the measurement from the RM analysis. 

\subsection{TOI-6029} \label{sec:toi6029_results}

\ptwo is the most massive planet in our sample, with $M_p=$ \mptwo $ M_J$, as well as the smallest ($R_p=$ \rptwo $ R_J$), indicating a high density relative to other hot Jupiters. Its host star is the hottest, most massive, and most luminous in the sample. TOI-6029 sits near the Kraft break at T$_{\rm eff} = 6223$ K, indicating that the system has had a substantial surface convective envelope ($\gtrsim$10\% of stellar radius) for only a relatively brief window of its lifetime. The out-of-transit residuals to our RV model have larger scatter than those of the other two systems, and we searched for additional non-transiting planets. We performed a Lomb-Scargle periodogram analysis on the residuals as well as attempting a two-planet fit to the observations, neither of which returned conclusive evidence of additional planetary companions. The lack of evidence for additional non-transiting planets indicates that the residual scatter is likely caused by stellar jitter. The spin-orbit alignment of the TOI-6029 system is similar to that measured for TOI-1181, with $\lambda =$ \lamptwo $^\circ$. The Doppler shadow measurement for \ptwo provides an obliquity constraint of $22.3\pm8.6^\circ$, which is consistent with the degree of misalignment provided by the RM measurement. We note that the sign of the obliquity from the different methods is flipped for \ptwo due to the degeneracy of the orientation of the transit chord.

\subsection{TOI-4379} \label{sec:toi4379_results}

\pthree is similar to Jupiter in mass ($M_p=$ \mpthree $ M_J$) but has a larger radius ($R_p=$ \rpthree $ R_J$). Its host star has the lowest projected stellar surface rotation velocity in our sample ($v\sin{i_\star} = 5.9\pm1.0$ km s$^{-1}$). The lack of sufficient RV baseline observations caused the measured sky-projected obliquity of the system to have larger uncertainty than the previously discussed measurements. The spin-orbit angle of the TOI-4379 system is consistent with being fully aligned ($\lambda =$ \lampthree $^\circ$).

\subsection{Population Comparison} \label{sec:pop_compare}

\begin{figure}[ht!]
    \centering
    \includegraphics[width=.45\textwidth]{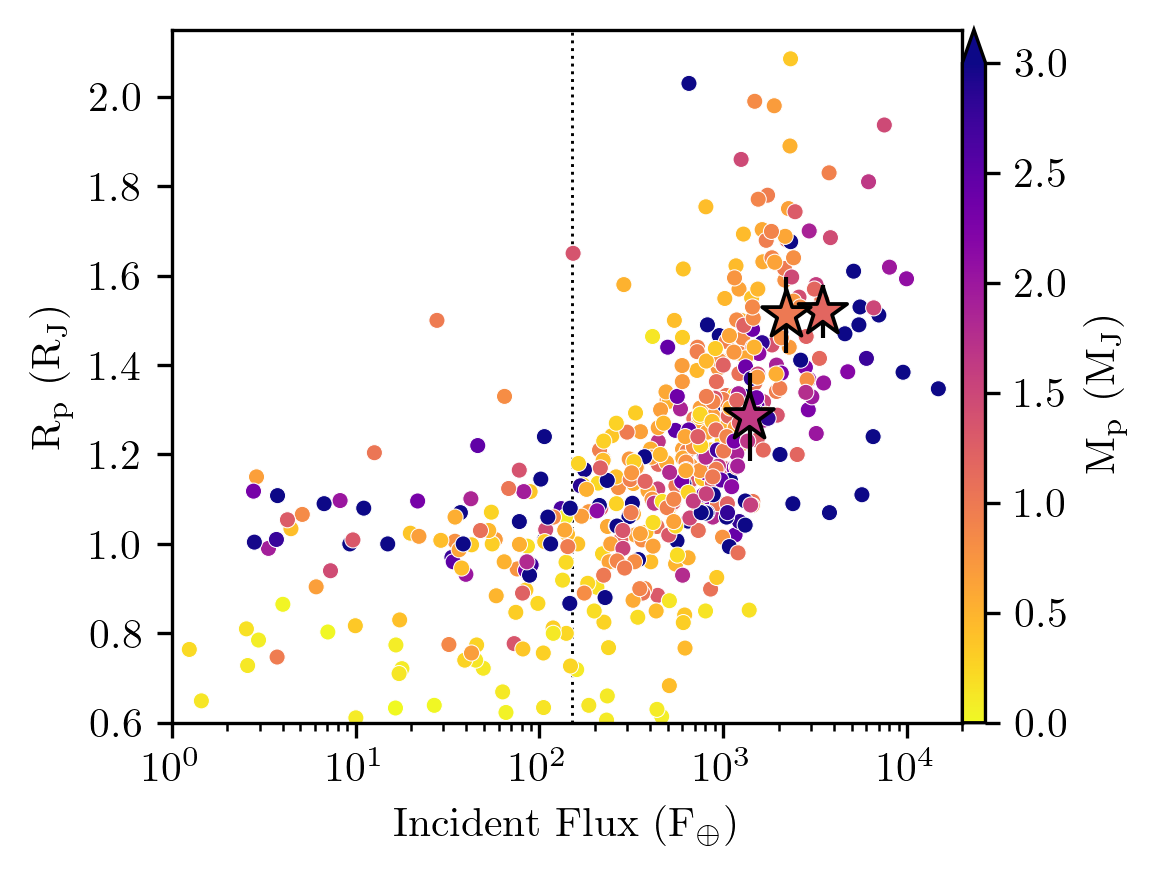}
\caption{Planet radius versus incident flux received for planets drawn from the literature (circles) and planets in our sample (stars). The colorbar indicates planet mass, and the dotted vertical line shows the empirical threshold above which planet inflation is observed.}
\label{fig:flux_radius}
\end{figure}

We now compare the radii, masses, and incident flux of our planet sample to the known population of exoplanets. Literature values were downloaded from the NASA Exoplanet Science Institute (NExScI) Exoplanet Archive\footnote{\href{https://exoplanetarchive.ipac.caltech.edu}{exoplanetarchive.ipac.caltech.edu}} on January 3, 2024. We limited the sample to exoplanets with both measured radii and masses, and removed planets for which the fractional uncertainty on either mass or radius exceeded 20\%. Figure \ref{fig:flux_radius} shows the well established relationship between incident flux, planet radius, and planet mass. We estimate incident flux with the relation from \cite{seager2010} given by
\[
\frac{F_{\rm p}}{F_\oplus} = \left(\frac{R_\star}{R_\oplus}\right)^2 \left(\frac{\rm T_{\rm eff,\star}}{\rm T_{\rm eff,\odot}}\right)^4 \left(\frac{a}{\rm AU}\right)^{-2}.
\]
\pone, \ptwo, and \pthree show inflation that is consistent with known planets of similar mass. The planets presented in this work constitute a small representative sample of hot Jupiters. We refer to \cite{chontos2024} for a population-level comparison of masses and radii for planets orbiting main-sequence and subgiant stars.



\section{Tidal Realignment} \label{sec:realign}

\subsection{Comparison to Main Sequence Obliquities} \label{sec:ms_compare}

\begin{figure}[h!]
    \centering
    \includegraphics[width=.5\textwidth]{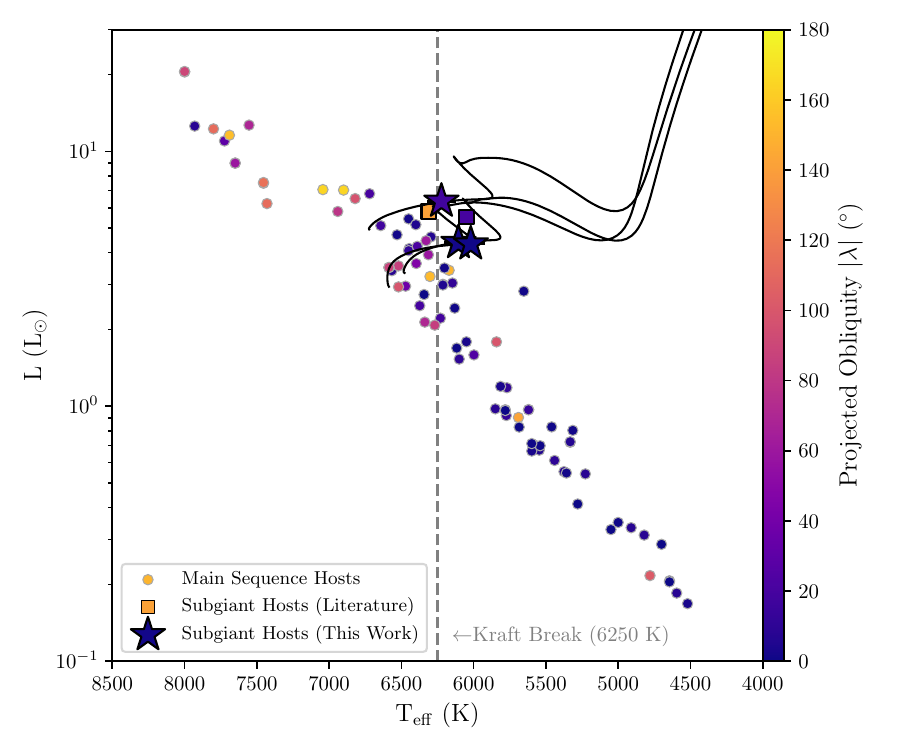}
    \caption{H-R diagram of stars hosting planets with measured sky-projected obliquities. Main sequence host stars are marked by circles, while subgiant hosts are shown as stars. The color of each point indicates the sky-projected obliquity, $\lambda$. Dotted black lines show our stellar evolution models fit to each of our host stars, and trace back to their ZAMS location.}
    \label{fig:lambda_hr}
\end{figure}

Figure \ref{fig:lambda_hr} shows a Hertzsprung-Russell (H-R) diagram of host stars for systems with measured sky-projected obliquities. To populate this figure, we downloaded obliquity measurements from the \textsf{TEPCat} catalog \citep{southworth11} and included additional measurements reported in \cite{albrecht2021}. We then performed cuts to select hot Jupiters systems ($M_P>0.3 M_J$, $P_{\rm orb} < 10$ days, $a/R_\star\leq8$) and removed obliquity measurements with uncertainties $>30$$^{\circ}$.


To estimate the evolution of our host stars, we fit evolutionary tracks from \textsf{MESA} Isochrones \& Stellar Tracks (\textsf{MIST}; \citealt{dotter16,choi16,paxton11,paxton13,paxton15}). We adopt the \textsf{isoclassify} mass and interpolate between values for metallicity in order to identify isochrone models that pass through our host stars near our stellar age estimates. The resulting tracks are shown as dotted lines in Figure \ref{fig:lambda_hr}. The evolutionary tracks show that these stars have indeed dropped below 6250 K during their post-main sequence evolution. In Figure \ref{fig:lambda_hr} we highlight two additional hot Jupiter systems for which the host star is likely somewhat evolved: WASP-71 b at 6180 K \citep{smith2013} and HAT-P-7 b at 6310 K \citep{winn2009}.

Near the estimated main-sequence positions of our targets, hot Jupiter hosts are observed with a wide range of obliquities. We test the likelihood that our three aligned systems are randomly drawn from the previously-aligned population in the observed distribution using a bootstrapping technique: we select a range of effective temperature between 6300 K and 6700 K which encompasses the zero-age main sequence (ZAMS) positions of our host stars. Next, we perform 100,000 random draws of three measured obliquities and calculate the fraction of draws in which all three systems were aligned ($<20$$^{\circ}$), and find that we retrieve the observed measurements $\sim$10\% of the time. If we include the literature measurement of WASP-71 b, the fraction drops to $\sim$4\%. This indicates that it is unlikely for our measured obliquities to be randomly sampled, however a larger sample would improve the confidence of this constraint.

Using a sample of subgiants which have decreased in temperature and crossed the Kraft break, we can test whether systems that were part of a frequently misaligned population become realigned after developing a deep surface convective envelope. Main sequence stars hotter than the Kraft break---similar to the H-R positions of our sample at the ZAMS---show a wide range of obliquities. If spin-orbit misalignment persists over the full lifetime of planetary systems, hot Jupiters orbiting stars which are hotter than the Kraft break should still be misaligned. In contrast, if the presence of a convective envelope causes spin-orbit realignment, the population of stars which gained deep convective envelopes after the main sequence should host hot Jupiters which are aligned or in the process of being aligned.


\subsection{Evolution of the Convective Envelope} \label{sec:conv_env}

Figure \ref{fig:lambda_hr} qualitatively suggests that misaligned hot Jupiters can be realigned as their host stars cross the Kraft break on the subgiant branch, implying efficient obliquity damping due to a convective envelope. To quantify whether obliquity damping can indeed act over subgiant evolution timescales, we examine the evolution of the surface convective envelope. We performed stellar modeling with the Modules for Experiments in Stellar Astrophysics code (\mesa; \citealt{paxton2010,paxton2013,paxton2015,paxton2018,paxton2019}) to trace the emergence of surface convective envelopes and estimate the time each star spent cooler than the Kraft break. Our models used initial elemental abundances from \cite{grevesse1998} and an atmospheric temperature structure following an Eddington $T(\uptau)$ relation with fixed opacity. We assume no diffusion and no core or envelope overshoot. 

\begin{figure*}[ht!]
    \centering
    \includegraphics[width=1\textwidth]{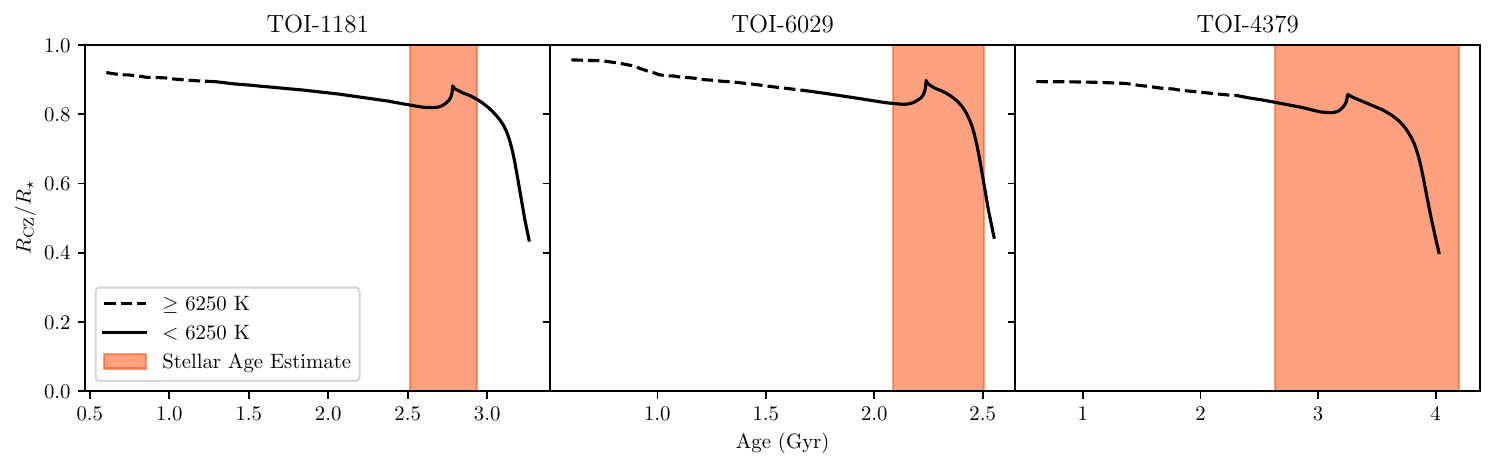}
    \caption{Fractional stellar radius, $R_{\rm CZ}/R_\star$, as a function of stellar age. On the y-axis, 0 represents the center of the star and 1 is its surface. The dashed line shows when the star was above the Kraft break ($\geq6250$ K) and the solid line shows when the star was below the Kraft break ($<6250$ K). The shaded regions show the range of estimated stellar ages from \textsf{isoclassify}.}
    \label{fig:r_cz}
\end{figure*}

We used the best-fit stellar properties for each star (given in Table \ref{tab:stellar}) as initial properties, and evolved our stellar models from the pre-main sequence until the base of the red giant branch. In Figure \ref{fig:r_cz}, we show the depth of the surface convection zone boundary in units of fractional stellar radius $R_{\rm CZ}/R_\star$ (where 0 is the center and 1 is the surface) as a function of stellar age.


Near the ZAMS, the convective boundary is near $R_{\rm CZ}/R_\star=1$, indicating a thin surface convection zone. As the stars evolve into subgiants, the convective boundary moves deeper into the stellar interior. While the mass fraction ($M_{\rm CZ}/M_\star$) contained in the convective envelope remains small ($\lesssim$0.01\%), the deepening surface convection zone may lead to enhanced tidal damping. We note that at the age when these stars drop below 6250 K, the outer $\sim$10\% of the stellar envelope is convective, but there is no clear feature to indicate a substantial deepening of the convective envelope over a short timescale. However, the observed change in the distribution of obliquities corresponds to this temperature, and we therefore select 6250 K as the threshold for enhanced damping. Realignment may require a minimum threshold of material for efficient damping, or it may correspond to the disappearance of the convective core. Further stellar modeling will be required to examine this threshold with more detail in the future.

If we assume that efficient realignment begins when a star decreases in temperature and drops below the Kraft break, we can roughly estimate the timescale for realignment. Because all systems in our sample are already aligned, we can only estimate an upper limit to this timescale by calculating the time between when the host star crossed the Kraft break and the estimated age at which it is observed today. This results in an upper limit on the realignment timescale of $\sim$1.5 Gyr for \pone, and $\sim$0.5 Gyr for \ptwo and \pthree. Using \ptwo and \pthree as references, we can place a global upper limit on the timescale for realignment at $\sim 500$ Myr. 

\subsection{Realignment Model} \label{sec:realign}

Stellar tides have been suggested to effectively damp out the obliquity of hot Jupiter systems, however they also damp orbital periods on similar or shorter timescales \citep{winn2010}. This leads us to an important question---how can a system realign without the planet facing runaway inspiral and engulfment? 

To test this, we model the evolution of orbital period and obliquity under the influence of tidal damping. We construct a toy model for orbital evolution, using a single stellar evolution model that is characteristic of stars in our sample, with $M_\star=1.4 M_\odot$, [Fe/H]$=0.3$ dex, and $M_p=1.2 M_J$. For simplicity, we assume a circular orbit for the planet. We use our toy model to explore the damping efficiencies required to reproduce a post-main sequence distribution similar to what we observe. 

We assume exponential decay prescribed by $\dot{a} = - a / \uptau_a$ for the semi-major axis, $a$, and $\dot{\lambda} = - \sin{(\lambda)} / \uptau_\lambda$ for the obliquity, $\lambda$.
The coefficients $\uptau_a$ and $\uptau_\lambda$ are the damping timescales, defined by \cite{lai2012} as
\begin{equation}
\uptau_a \approx 1.28 \left( \frac{Q'_\star}{10^7}\right) \left(\frac{M_\star}{10^3 M_{\rm p}} \right) \left(\frac{\bar{\rho}_\star}{\bar{\rho}_\odot}\right)^{5/3} \left(\frac{P_{\rm orb}}{1{\rm d}}\right)^{13/3} {\rm Gyr}   
\label{eq1}
\end{equation}
and 
\begin{equation}
\uptau_\lambda \approx 1.13 \uptau_a \left(\frac{\kappa}{0.1}\right) \left(\frac{M_\star}{10^3 M_{\rm p}}\right) \left(\frac{\bar{\rho}_\odot}{\bar{\rho}_\star}\right)^{2/3} \left(\frac{10{\rm d}}{P_{\rm s}}\right) \left(\frac{1{\rm d}}{P_{\rm orb}}\right)^{1/3}    
\label{eq2}
\end{equation}
where $Q'_\star$ is the tidal quality factor of the star, $\bar{\rho}_\star$ is the mean stellar density, and $\kappa$ is related to the stellar moment of inertia $I_\star$ by $I_\star = \kappa M R^2$.

The calculation of $\uptau_\lambda$ in Equation \ref{eq2} requires knowledge of the stellar rotation period. We therefore produced our stellar models with a modified version of \mesa which calculates and stores parameters necessary to trace the rotational evolution of the star (see \citealt{saunders2024} for details). We assume standard Skumanich-like spindown over the main sequence lifetime. This is a reasonable assumption for stars in this mass range, as they are not expected to undergo significant magnetic braking \citep{vansaders16,vansaders19}. 

\begin{figure*}[ht!]
\centering
\gridline{\fig{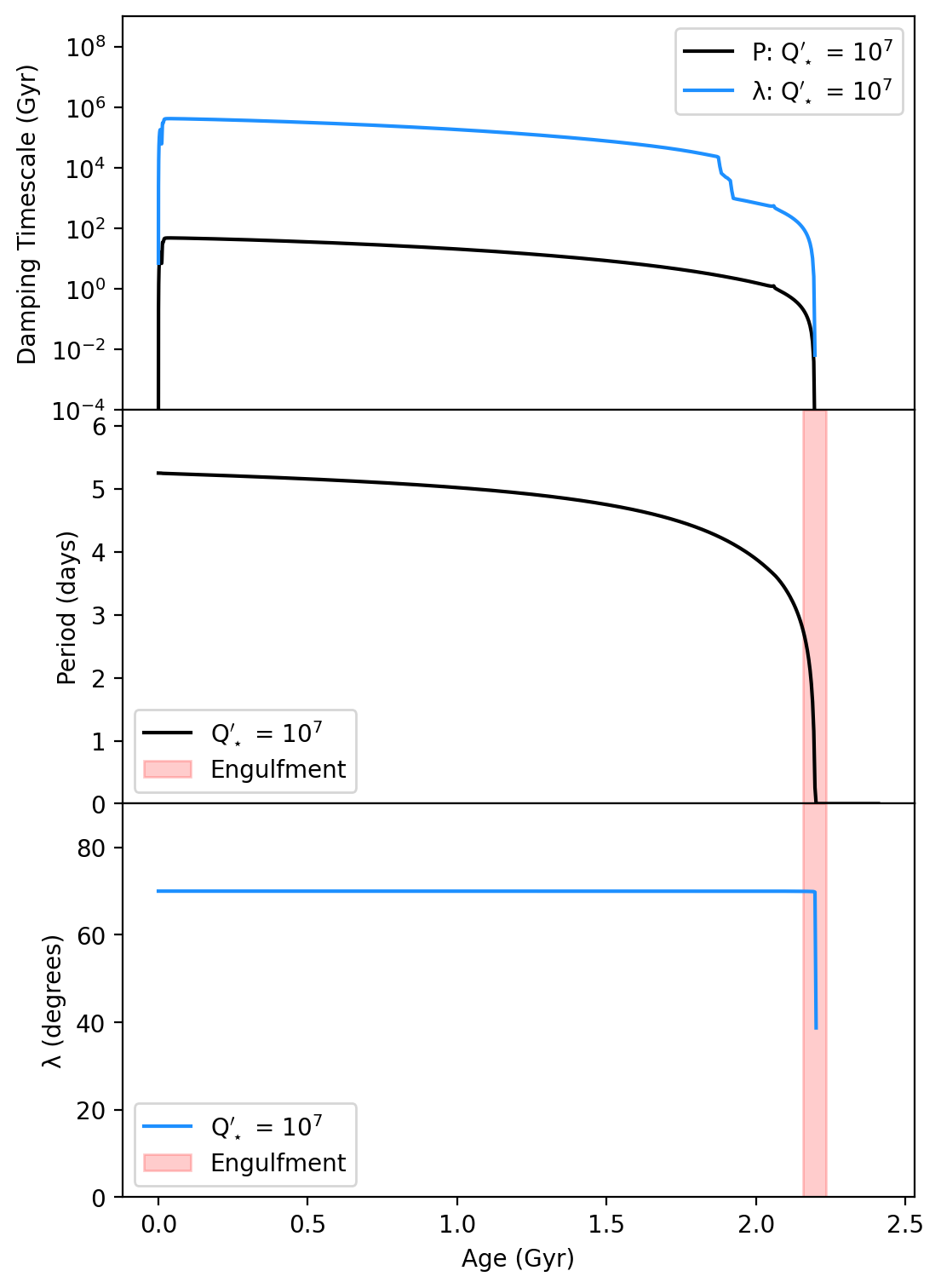}{.45\textwidth}{\textbf{(a)}}
          \fig{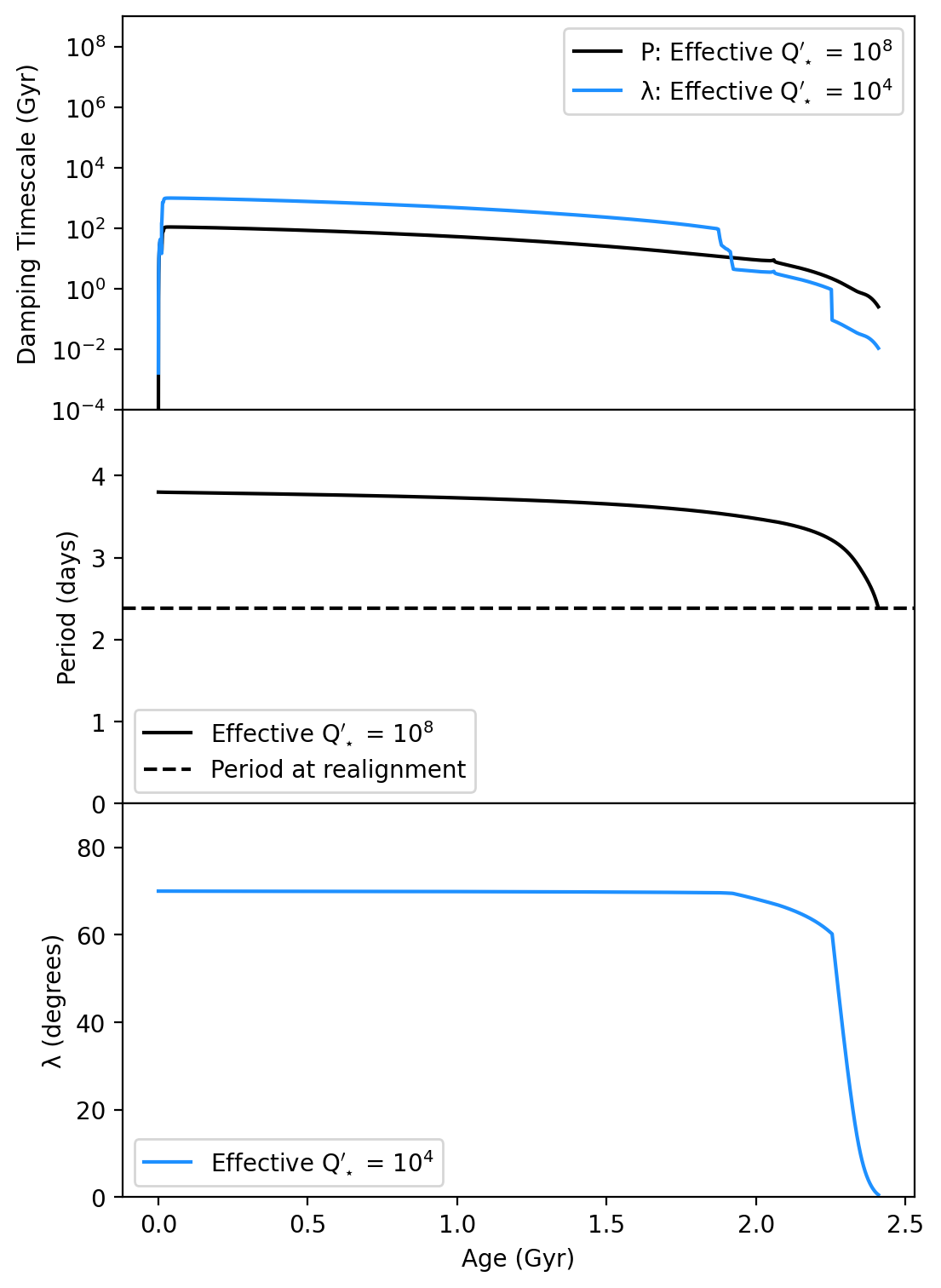}{.45\textwidth}{\textbf{(b)}}}
\caption{\textbf{(a)} \textit{Top:} Damping timescales for our toy model, calculated using Equations 1 \& 2, as a function of stellar age, with $\uptau_a$ in black and $\uptau_\lambda$ in blue. \textit{Middle:} Model for the evolution of orbital period, $P$ (shown by the black line). \textit{Bottom:} Model for the evolution of obliquity, $\lambda$ (shown by the blue line). These models were computed assuming the damping efficiency is the same for both parameters. The red region indicates the runaway inspiral regime. \textbf{(b)} Same as panel (a), but with different efficiencies for the damping of $P$ and $\lambda$, prescribed by the tidal quality factor $Q'_\star$. This model has a difference in effective tidal quality factor of $\delta\log{Q'_\star}=4$.}
\label{fig:damping}
\end{figure*}

The choice of tidal quality factor $Q'_\star$ has a significant effect on the dynamical evolution. However, values for $Q'_\star$ range over multiple orders of magnitude for subgiants \citep{patel2022}. In Figure \ref{fig:damping}, we show the impact of this choice on our dynamical models. If $Q'_\star$ is assumed to be the same for the damping of $P$ and $\lambda$, the planet enters a phase of runaway inspiral and is engulfed before it can realign (panel a). If we instead define the damping timescales $\uptau_a$ and $\uptau_\lambda$ with different \textit{effective} tidal quality factors, it is possible to produce a model in which the planet realigns before it can be engulfed. Because the obliquity distribution appears distinct on either side of the Kraft break, we assume that $\uptau_a$ and $\uptau_\lambda$ share the same tidal quality factor until the model drops below 6250 K, at which point we invoke a deviation in $Q'_\star$, described by the order of magnitude difference $\delta\log{Q'_\star}$. Figure \ref{fig:damping}b shows that a difference in effective tidal quality factor $\delta\log{Q'_\star}\gtrsim4$ is required to efficiently realign hot Jupiters on orbits between $\sim$3-8 days without entering a phase of runaway inspiral. 

\begin{figure*}[ht!]
    \centering
    \includegraphics[width=1\textwidth]{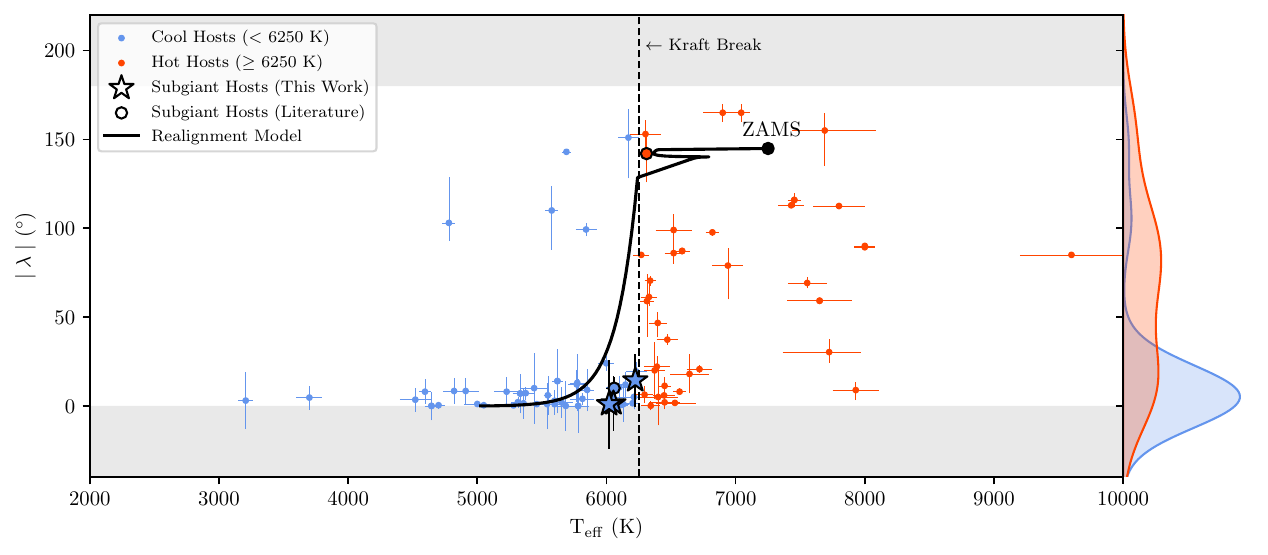}
    \caption{Distribution of hot Jupiter obliquities as a function of stellar effective temperature, T$_{\rm eff}$. Blue points indicate cool ($<6250$ K) hosts, and red points indicate hot ($>6250$ K) hosts. The vertical dashed black line marks the Kraft break. Subgiant hosts are marked by large circles with black outlines, and all other points are main sequence hosts. The distributions on the far right of the plot show the kernel densities of each sample, corresponding to color. The black line shows a model for the post-main sequence realignment of a hot Jupiter. This model shows a possible pathway from the hot, misaligned population to the observed cool, well-aligned population.}
    \label{fig:realignment}
\end{figure*}

Figure \ref{fig:realignment} shows the sample from Figure \ref{fig:lambda_hr} but instead plotting obliquity versus effective temperature, highlighting the correlation between alignment and stellar effective temperature. To examine the change in $\lambda$ as the host star evolves, we take the model in Figure \ref{fig:damping} b and plot it in $\lambda$ versus T$_{\rm eff}$ space. We observe that the toy model indeed reproduces an evolutionary track in which a deepening convective envelope during subgiant evolution causes obliquity damping.

The toy model shown in Figures \ref{fig:damping} and \ref{fig:realignment} make several assumptions. First, we assume that the tidal quality factor $Q'_\star$ does not evolve smoothly in time, and instead drops precipitously in the case of $\lambda$-damping after the star crosses the Kraft break. It is unlikely that this fully captures the behavior seen in nature, as $Q'_\star$ depends on the interior stellar structure, which evolves smoothly over main sequence lifetimes. However, main sequence internal structure changes are small, and we choose to fix the values of $Q'_\star$ on the main sequence, as well as after the threshold is crossed. We also assume that the value of $Q'_\star$ applied to the damping of period is held fixed for the entire evolution of the star. As the stellar structure evolves after the star leaves the main sequence, its $Q'_\star$ should change, though the relationship between $Q'_\star$ and stellar radius and density is uncertain. We again choose the simple case of holding the value fixed, though future studies of how $Q'_\star$ relates to stellar interior evolution may allow us to improve this model in the future. There are promising pathways to better characterize $Q'_\star$ through eclipsing binary system evolution \citep{fabrycky07} and asteroseismology \citep{chontos2019}.


We now consider what physical mechanisms could lead to a large difference in the efficiencies of the orbital period and obliquity damping timescales. \cite{winn10} invoke sustained core-envelope decoupling as a potential source of enhanced obliquity damping. In this scenario, the planet is able to exert a torque on the outer envelope of the star without rapidly losing orbital angular momentum. They note, however, that this explanation is challenged by evidence that the core and envelope of main sequence stars appear to be well-coupled \citep{winn10,jianke2013}. It is unclear whether this expectation applies to subgiant stars, and the theory of sustained core-envelope decoupling could be tested with asteroseismic measurements of radial differential rotation or core-envelope misalignments in evolved stars \citep{lund14}.

Another explanation, posed by \cite{lai2012}, suggests that in the presence of a convective envelope, a hot Jupiter on a misaligned orbit can excite inertial waves which dissipate angular momentum as they propagate through the stellar interior. \cite{ogilvie2007} examined tidal dissipation in stars with convective envelopes in the context of hydrodynamical theory, and found that these inertial waves can result in the value of $Q'_\star$ being reduced by up to 4 orders of magnitude. Because these inertial waves are static in the inertial frame, they do not contribute to orbital decay \citep{lai2012}, and only impact the efficiency of obliquity damping. This suggested order of magnitude difference is consistent with our toy model in Figures \ref{fig:damping} and \ref{fig:realignment} following the prescription for tidal damping laid out in \cite{lai2012}.

A recent study by \cite{zanazzi2024} showed that hot Jupiter orbits can resonantly lock to gravity mode pulsations in the radiative interior of cool stars and efficiently damp the system's obliquity (similar to semi-major axis damping described in \citealt{ma2021}). This mechanism is able to damp the stellar obliquity without causing the planet to inspiral into its host star. The interior structure of our subgiant hosts is amenable to gravity mode pulsations, providing another possible mechanism for the observed obliquity damping.


\subsection{Predicting Orbital Evolution of Main Sequence Systems} \label{sec:implications}

To investigate period and obliquity damping over a range of parameters, we calculate our toy model over a grid of main sequence orbital period, $P_{\rm ZAMS}$, planet mass, $M_{\rm p}$, and $\delta\log{Q'_\star}$. We consider a system ``engulfed'' if the orbital period is reduced to zero before the star reaches the base of the red giant branch, ``realigned'' if the obliquity drops below 10$^\circ$ and the orbital period is greater than zero for the full track, and ``not aligned'' if the planet is not engulfed but retains an obliquity greater than 10$^\circ$. This threshold corresponds to the typical uncertainty on a sky-projected obliquity measurement (see Figure \ref{fig:realignment}).

\begin{figure}[ht!]
    \includegraphics[width=.5\textwidth]{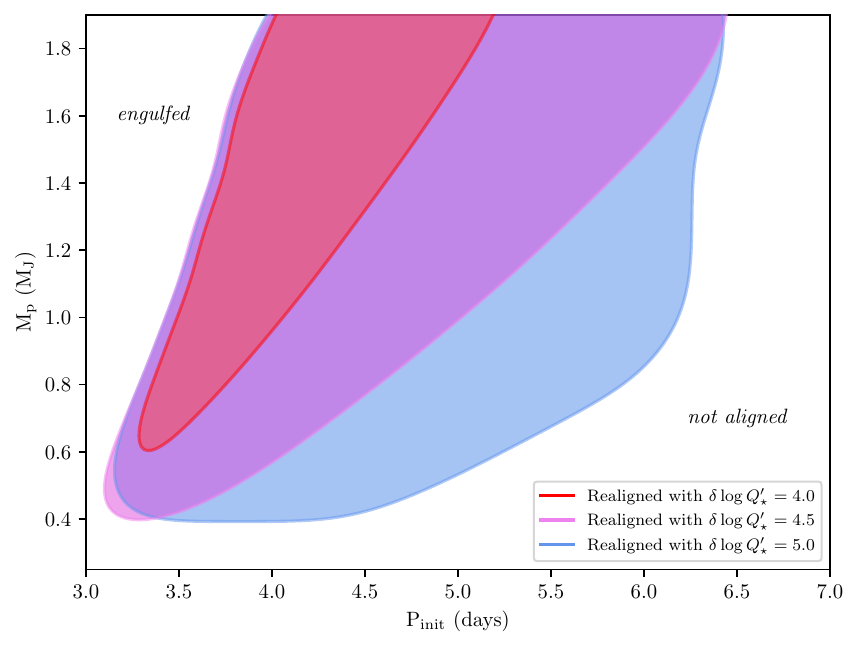}
    \caption{Regions of expected realignment given different values of $\delta\log{Q'_\star}$. The x-axis shows the initial orbital period of the planet (at ZAMS), and the y-axis shows the planet mass. Inside the shaded contours, realignment without engulfment is expected. Below the contours, $\lambda$ is not damped efficiently enough to realign, and above the contours the planet enters runaway inspiral.}
    \label{fig:contours}
\end{figure}

Figure \ref{fig:contours} shows the parameter space where realignment is expected for a range of values for $\delta\log{Q'_\star}$ as a function of main sequence orbital period, $P_{\rm ZAMS}$, and planet mass, $M_{\rm p}$. The regions inside the contours are expected to realign without being engulfed before the star reaches the base of the red giant branch. As the orbital period increases, a larger $M_{\rm p}$ is required to efficiently damp the obliquity. However, if the planet mass is too high, the damping of orbital period will be efficient enough to catalyze runaway inspiral. Figure \ref{fig:contours} can be used as a rough guide to predict whether a hot Jupiter orbiting a hot main sequence star will realign after the main sequence.

\section{Conclusions}\label{sec:conclusions}

The obliquities of hot Jupiter systems have been observed to show a strong dependence on the effective temperature of the host star, with cool stars ($<6250$ K) hosting mostly aligned planets and hot stars ($\geq 6250$ K) hosting mostly misaligned planets. Theories have suggested that a surface convective envelope can damp the obliquity of a hot Jupiter's orbit.

To test this theory, we have measured the sky-projected spin-orbit angles for three subgiants hosting hot Jupiters. Each of the host stars in our sample crossed the Kraft break and gained deep surface convective envelopes during their post-main sequence evolution. Our evolved hot Jupiter systems have low sky-projected obliquities, indicating efficient realignment as stars develop convective envelopes. 

We summarize our main conclusions as follows:
\begin{enumerate}
    \item We confirm two newly identified hot Jupiters orbiting subgiant stars: \ptwo ($R_{\rm p} =$ \rptwo $R_{\rm J}$, $M_{\rm p} = $ \mptwo $ M_{\rm J}$, $P_{\rm orb} = $ \porbtwo days), and \pthree ($R_{\rm p} = $ \rpthree $ R_{\rm J}$, $M_{\rm p} = $ \mpthree $ M_{\rm J}$, $P_{\rm orb} =$ \porbthree days).
    \item We measured the RM effect for three hot Jupiters orbiting subgiants that have recently developed convective envelopes. All three are aligned ($|\lambda| < 15^\circ$), which is consistent with the picture that convective envelopes can efficiently damp obliquities. We place an upper limit on the timescale for tidal realignment of $\sim$500 Myr.
    \item A simple model incorporating tidal realignment to explain these observations requires the effective tidal quality factor for obliquity damping to be at least four orders of magnitude lower than that for orbital decay ($\delta\log Q'_\star\gtrsim 4$). This is consistent with angular momentum dissipation through the driving and damping of inertial waves in the stellar convective envelope or coupling to gravity modes in the radiative interior.
\end{enumerate}

Further work is required to place tighter constraints on the timescales for obliquity damping. A larger sample of obliquity measurements for planets orbiting subgiant stars similar to those reported here would provide additional benchmarks for the dynamical evolution, and potentially distinguish between mechanisms for enhanced $\lambda$-damping efficiency. One approach to better understand the relationship between $\lambda$ and T$_{\rm eff}$ would be to measure the stellar obliquity for massive subgiants which have not yet crossed the Kraft break. By observing systems at a variety of evolutionary stages, we will better constrain the efficiency of obliquity damping to test high-eccentricity migration as the mechanism for hot Jupiter formation.

\section*{Acknowledgements}\label{sec:acknowledgements}

N.S. acknowledges support by the National Science Foundation Graduate Research Fellowship Program under Grant Numbers 1842402 \& 2236415. A.C. \& D.H. acknowledge support from the National Aeronautics and Space Administration (80NSSC21K0652). D.H. acknowledges support from the Alfred P. Sloan Foundation and the Australian Research Council (FT200100871). M.R. is supported by Heising-Simons grant \#2023-4478. K.A.C. and C.N.W. acknowledge support from the \tess mission via subaward s3449 from MIT. T.D. acknowledges support by the McDonnell Center for the Space Sciences at Washington University in St. Louis. J.M.A.M. is supported by the National Science Foundation (NSF) Graduate Research Fellowship Program (GRFP) under Grant No. DGE-1842400. 


This paper made use of data collected by the \tess mission and are publicly available from the Mikulski Archive for Space Telescopes (MAST) operated by the Space Telescope Science Institute (STScI). Funding for the \tess mission is provided by NASA's Science Mission Directorate. We acknowledge the use of public \tess data from pipelines at the \tess Science Office and at the \tess Science Processing Operations Center. Resources supporting this work were provided by the NASA High-End Computing (HEC) Program through the NASA Advanced Supercomputing (NAS) Division at Ames Research Center for the production of the SPOC data products. Funding for the \tess mission is provided by NASA's Science Mission Directorate. 

Some of the data presented herein were obtained at Keck Observatory, which is a private 501(c)3 non-profit organization operated as a scientific partnership among the California Institute of Technology, the University of California, and the National Aeronautics and Space Administration. The Observatory was made possible by the generous financial support of the W. M. Keck Foundation. 

The authors wish to recognize and acknowledge the very significant cultural role and reverence that the summit of Maunakea has always had within the Native Hawaiian community. We are most fortunate to have the opportunity to conduct observations from this mountain.


This work makes use of observations from the LCOGT network. Part of the LCOGT telescope time was granted by NOIRLab through the Mid-Scale Innovations Program (MSIP). MSIP is funded by NSF.

Some of the observations in this paper made use of the High-Resolution Imaging instrument Zorro and were obtained under Gemini LLP Proposal Number: GN/S-2021A-LP-105. Zorro was funded by the NASA Exoplanet Exploration Program and built at the NASA Ames Research Center by Steve B. Howell, Nic Scott, Elliott P. Horch, and Emmett Quigley. Zorro was mounted on the Gemini South telescope of the international Gemini Observatory, a program of NSF's OIR Lab, which is managed by the Association of Universities for Research in Astronomy (AURA) under a cooperative agreement with the National Science Foundation. on behalf of the Gemini partnership: the National Science Foundation (United States), National Research Council (Canada), Agencia Nacional de Investigación y Desarrollo (Chile), Ministerio de Ciencia, Tecnología e Innovación (Argentina), Ministério da Ciência, Tecnologia, Inovações e Comunicações (Brazil), and Korea Astronomy and Space Science Institute (Republic of Korea).


This research has made use of the Exoplanet Follow-up Observation Program (ExoFOP; DOI: 10.26134/ExoFOP5) website, which is operated by the California Institute of Technology, under contract with the National Aeronautics and Space Administration under the Exoplanet Exploration Program.

\software{\textsf{Lightkurve} \citep{lightkurve}, \textsf{Astropy} \citep{astropy2013,astropy2018,astropy2022}, \textsf{Astroquery} \citep{astroquery}}

\pagebreak
\nocite{tange2018}
\bibliography{references,references2}{}
\bibliographystyle{aasjournal}

\appendix 

\section{Ground-based Imaging Figures} \label{sec:imaging_figures}

\begin{figure}[h!]
    \centering
    \includegraphics[width=.5\textwidth]{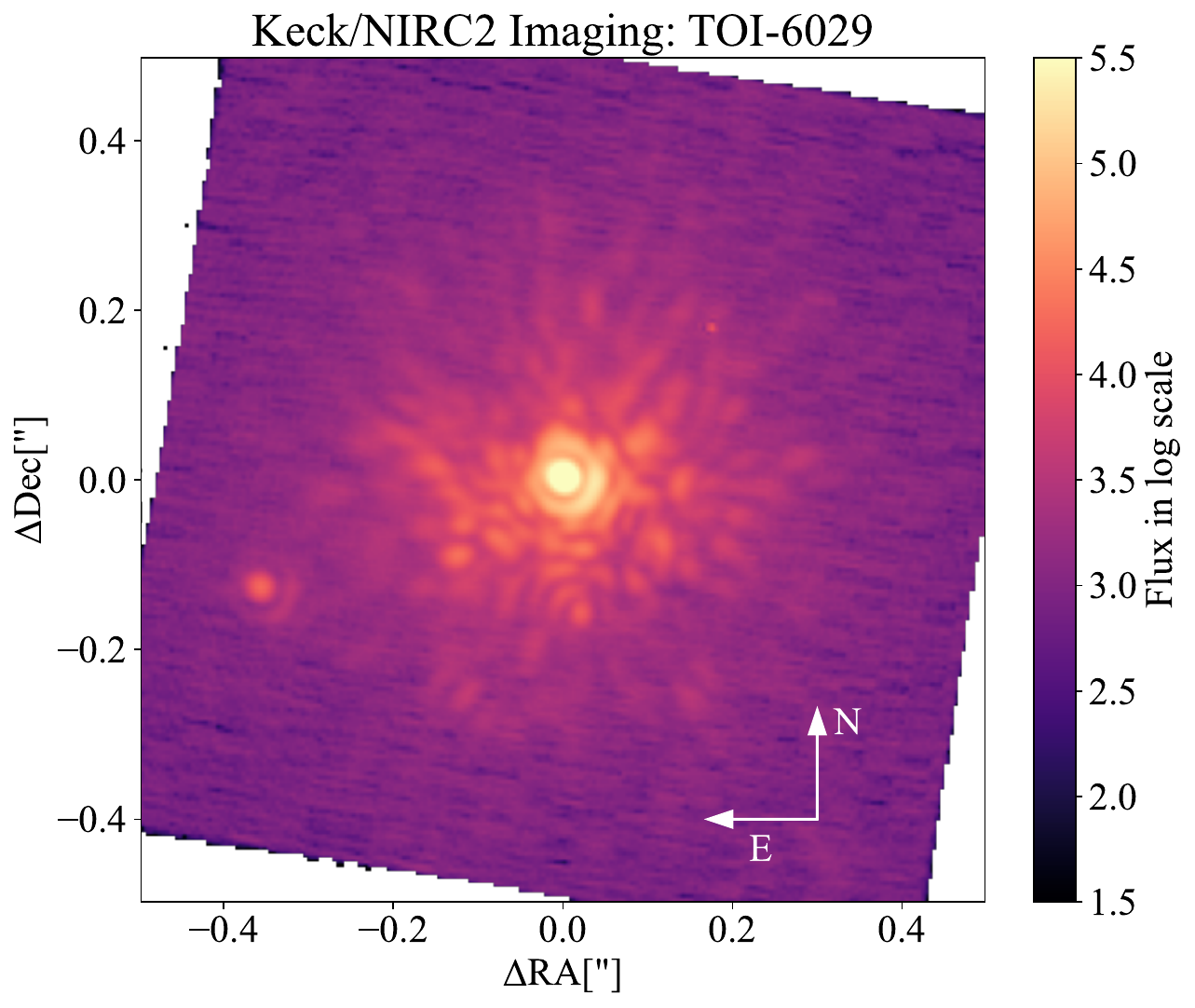}
    \caption{High-contrast image of \stwo taken with the Keck/NIRC2 near-infrared imager in the Kp bandpass. The image has been corrected and derotated. The colorbar indicates the measured flux.}
    \label{fig:toi6029_image}
\end{figure}

Figure \ref{fig:toi6029_image} shows the high-contrast image of \stwo obtained with the Keck/NIRC2 near-infrared imager (for details, see \S \ref{sec:hci}). The corrected and derotated image, taken in the Kp bandpass, shows a companion with a separation of $763\pm20$ milliarcseconds at a position angle of $110.2\pm0.5^{\circ}$.

\begin{figure*}[ht!]
\centering
\gridline{\fig{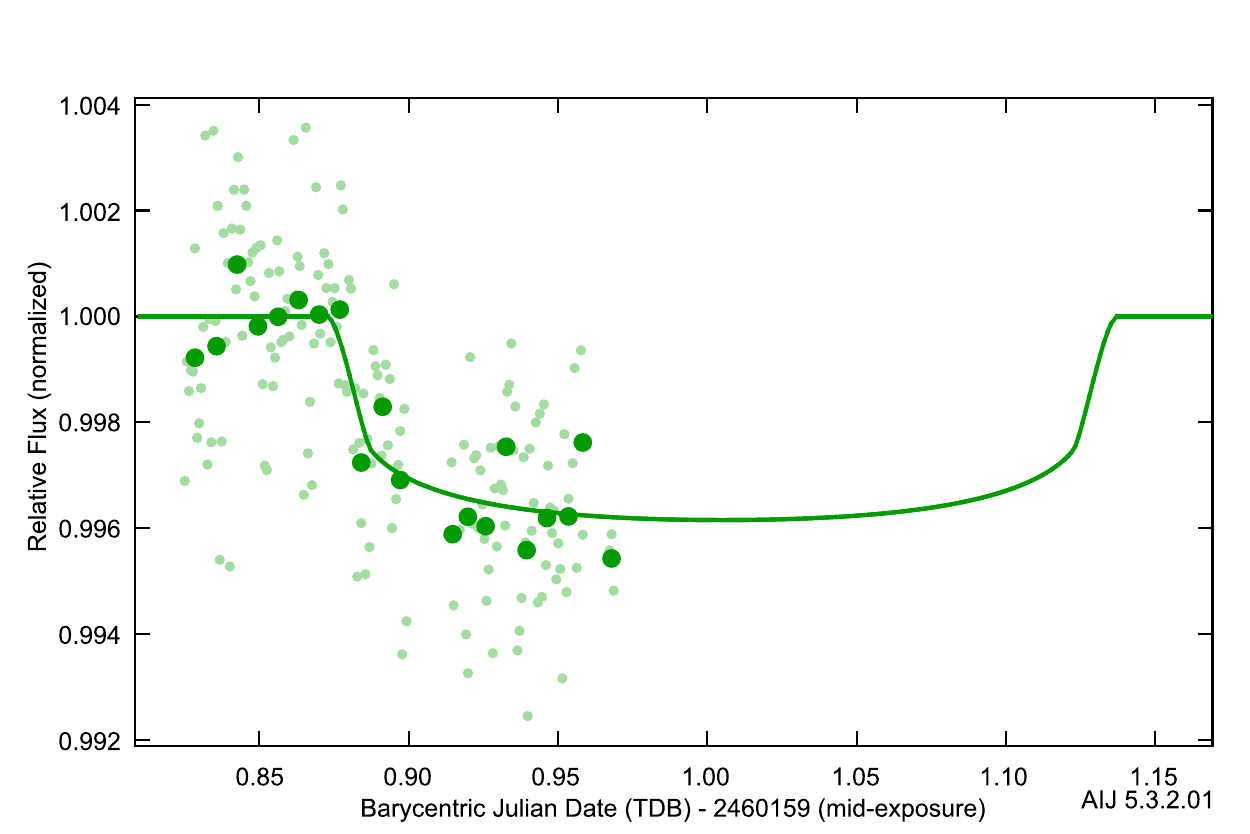}{.33\textwidth}{(a)}
          \fig{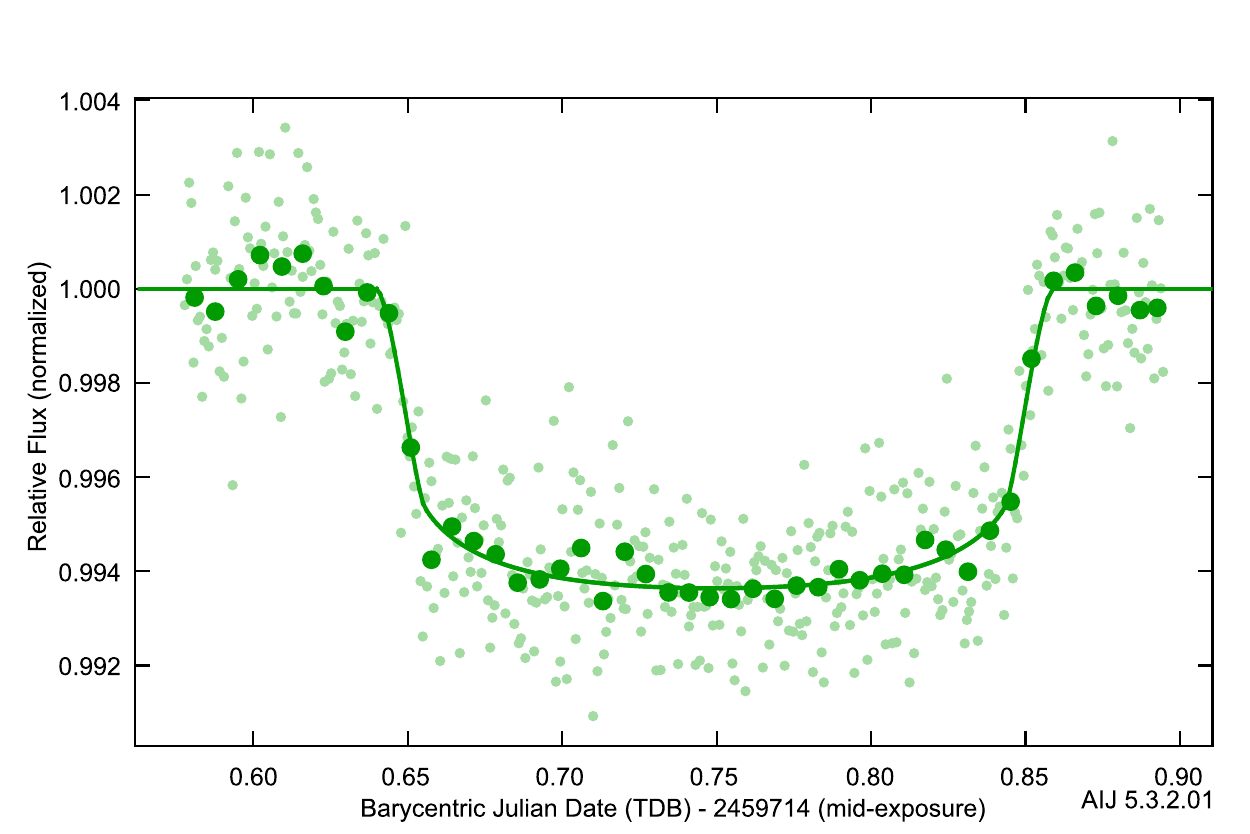}{.33\textwidth}{(b)}
          \fig{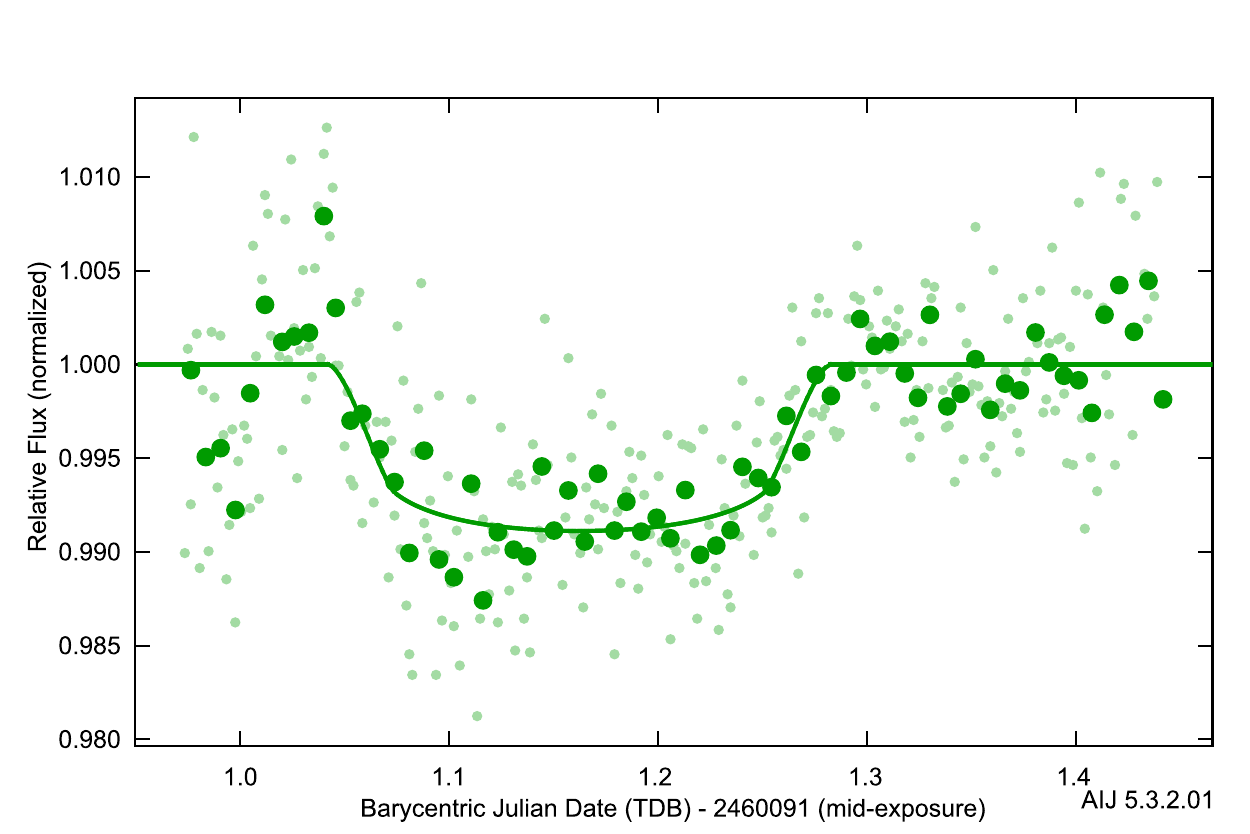}{.33\textwidth}{(c)}}
\caption{Ground-based time-series photometry observations of the transits of \ptwo and \pthree. The light green symbols show the unbinned data. The larger dark green symbols show the data binned in 10 minute bins. The solid green line is the best fit transit model. \textbf{(a)} Partial transit of \ptwo observed by LCOGT on August 3, 2023. \textbf{(b)} Transit of \pthree observed by LCOGT on May 15, 2022. \textbf{(c)} Transit of \pthree observed by PEST on May 27, 2023.}
\label{fig:ground_phot}
\end{figure*}

Figure \ref{fig:ground_phot} shows photometric time-series observations of the transits of \ptwo and \pthree. These data were obtained with the ground-based LCOGT and PEST telescopes (for details, see \S \ref{subsec:ground}), and they display transit depths and ephemerides consistent with our predictions. 

\section{Radial Velocity Measurements} \label{sec:rv_obs}

Table \ref{tab:my_label} lists RV measurements obtained with Keck/HIRES and Keck/KPF for each of our targets. A machine-readable version of this table containing all observations is available.

\begin{footnotesize}
\begin{longtable}{l l l l l}
    \toprule
    Target & Instrument & Time (BJD) & RV (m/s) & RV Uncertainty (m/s) \\ 
    \midrule
    TOI-1181 & Keck/HIRES & $2459000.040402$ & $-67.993$ & $3.735$ \\
TOI-1181 & Keck/HIRES & $2459003.045446$ & $14.927$ & $3.907$ \\
TOI-1181 & Keck/HIRES & $2459004.050001$ & $72.699$ & $3.539$ \\
TOI-1181 & Keck/HIRES & $2459026.093511$ & $-50.794$ & $4.016$ \\
TOI-1181 & Keck/HIRES & $2459029.024667$ & $118.585$ & $3.540$ \\
\vdots & & & & \\
\midrule
TOI-6029 & Keck/HIRES & $2459423.093788$ & $84.436$ & $6.597$ \\
TOI-6029 & Keck/HIRES & $2459442.102513$ & $-136.860$ & $6.772$ \\
TOI-6029 & Keck/HIRES & $2459447.07665$ & $-26.309$ & $7.122$ \\
TOI-6029 & Keck/HIRES & $2459449.108516$ & $-30.814$ & $6.492$ \\
TOI-6029 & Keck/HIRES & $2459457.094269$ & $111.598$ & $6.563$ \\
\vdots & & & & \\
\midrule
TOI-4379 & Keck/HIRES & $2459478.753368$ & $-38.193$ & $8.845$ \\
TOI-4379 & Keck/HIRES & $2459482.716898$ & $104.904$ & $5.204$ \\
TOI-4379 & Keck/HIRES & $2459633.119397$ & $63.349$ & $4.963$ \\
TOI-4379 & Keck/HIRES & $2459694.041773$ & $79.915$ & $5.086$ \\
TOI-4379 & Keck/HIRES & $2459710.004739$ & $24.236$ & $4.605$ \\
\vdots & & & & \\
TOI-4379 & Keck/KPF & $2460101.830527187$ & $68.210$ & $15.086$ \\
TOI-4379 & Keck/KPF & $2460101.837867083$ & $49.998$ & $9.299$ \\
TOI-4379 & Keck/KPF & $2460101.844402326$ & $37.009$ & $7.578$ \\
TOI-4379 & Keck/KPF & $2460101.85159022$ & $40.815$ & $9.384$ \\
TOI-4379 & Keck/KPF & $2460101.858644097$ & $48.221$ & $7.549$ \\
\vdots & & & & \\
\bottomrule
    \caption{Radial velocity measurements for TOI-1181, TOI-6029, and TOI-4379 used in our analysis. A full machine-readable table can be accessed online.}
    \label{tab:my_label}
\end{longtable}
\end{footnotesize}

\end{CJK*}
\end{document}